\documentclass[12pt]{article}
\usepackage{amsmath,amssymb}
\usepackage{graphicx}
\newcommand{\eqdef}{\stackrel{\text{def}}{=}}
\newcommand{\n}{\nonumber\\}
\newcommand{\bm}{\boldsymbol}
\newcommand{\cF}{c_{\text{\tiny$\mathcal{F}$}}}
\newcommand{\ignore}[1]{}
\numberwithin{equation}{section}
\newcommand{\Romannumeral}[1]{\uppercase\expandafter{\romannumeral#1}}
\newcommand{\I}{\text{\Romannumeral{1}}}
\newcommand{\II}{\text{\Romannumeral{2}}}
\newcommand{\III}{\text{\Romannumeral{3}}}
\newtheorem{conj}{\bf Conjecture}
\newtheorem{prop}{\bf Proposition}
\newtheorem{lemma}{\bf Lemma}

\allowdisplaybreaks[4]

\begin{document}

\baselineskip=20pt

\newfont{\elevenmib}{cmmib10 scaled\magstep1}
\newcommand{\preprint}{
    \begin{flushright}\normalsize \sf
     DPSU-14-3\\
   \end{flushright}}
\newcommand{\Title}[1]{{\baselineskip=26pt
   \begin{center} \Large \bf #1 \\ \ \\ \end{center}}}
\newcommand{\Author}{\begin{center}
   \large \bf Satoru Odake \end{center}}
\newcommand{\Address}{\begin{center}
     Department of Physics, Shinshu University,\\
     Matsumoto 390-8621, Japan
   \end{center}}
\newcommand{\Accepted}[1]{\begin{center}
   {\large \sf #1}\\ \vspace{1mm}{\small \sf Accepted for Publication}
   \end{center}}

\preprint
\thispagestyle{empty}

\Title{Recurrence Relations of\\ the Multi-Indexed Orthogonal Polynomials
: $\II$}

\Author

\Address
\vspace{1cm}

\begin{abstract}
In a previous paper we presented $3+2M$ term recurrence relations with
variable dependent coefficients for $M$-indexed orthogonal polynomials of
Laguerre, Jacobi, Wilson and Askey-Wilson types.
In this paper we present (conjectures of) the recurrence relations with
constant coefficients for these multi-indexed orthogonal polynomials.
The simplest recurrence relations have $3+2\ell$ terms, where $\ell\,(\geq M)$
is the degree of the lowest member of the multi-indexed orthogonal polynomials.
\end{abstract}

\section{Introduction}
\label{intro}

Exactly solvable quantum mechanical systems in one dimension have seen
remarkable developments in recent years and the central role is played by
exceptional orthogonal polynomials \cite{gomez}--\cite{equiv_miop}(and
the references therein).
A set of polynomials $\{\mathcal{P}_n(\eta)|n\in\mathbb{Z}_{\geq 0}\}$ is
called exceptional orthogonal polynomials, when the following conditions
(\romannumeral1)--(\romannumeral3) plus (\romannumeral4) are satisfied;
(\romannumeral1) they are orthogonal with respect to some inner product,
(\romannumeral2) there are missing degrees, i.e.,
$\{\text{deg}\,\mathcal{P}_n|n\in\mathbb{Z}_{\geq 0}\}\subsetneqq
\mathbb{Z}_{\geq 0}$,
(\romannumeral3) but they form a complete set, and
(\romannumeral4) they satisfy second order differential or difference equations.
The constraints of Bochner's theorem and its generalizations
\cite{bochner,szego} are avoided by the condition (\romannumeral2).
We want to distinguish the following two cases;
the set of missing degrees $\mathcal{I}=\mathbb{Z}_{\geq 0}\backslash
\{\text{deg}\,\mathcal{P}_n|n\in\mathbb{Z}_{\geq 0}\}$ is
case (1): $\mathcal{I}=\{0,1,\ldots,\ell-1\}$,
case (2) $\mathcal{I}\neq\{0,1,\ldots,\ell-1\}$, where $\ell$ is a positive
integer. The situation of case (1) is called stable in \cite{gomez12}.
The first example of the case (1) exceptional orthogonal polynomials,
$X_1$ Laguerre
and Jacobi, was found by G\'{o}mez-Ullate, Kamran and Milson \cite{gomez},
and its quantum mechanical formulation was given by Quesne \cite{quesne}.
Based on the quantum mechanical formulation (ordinary quantum mechanics (oQM),
discrete quantum mechanics with pure imaginary shifts (idQM)\cite{os24}),
Sasaki and the present author constructed $X_{\ell}$ polynomials and their
generalizations, multi-indexed orthogonal polynomials
\cite{os16,os17,os25,os27}.
The multi-indexed orthogonal polynomials of Laguerre, Jacobi, Wilson and
Askey-Wilson types, which are obtained by multi-step Darboux transformations
\cite{darb,crum,adler,os15,gos}
with virtual state wavefunctions as seed solutions \cite{os25,os27},
correspond to the case (1).
The exceptional orthogonal polynomials, which are obtained by multi-step
Darboux transformations with eigenstate or pseudo virtual state wavefunctions
as seed solutions \cite{os29,os28,gos,os30}, correspond to the case (2).
For those having purely discrete orthogonality weight functions,
the number of orthogonal polynomials may be
finite. The multi-indexed ($q$-)Racah polynomials \cite{os23,os26},
which are constructed based on discrete quantum mechanics with real shifts
(rdQM) \cite{os24}, correspond to the case (1).

The ordinary orthogonal polynomials
$\{P_n(\eta)|n\in\mathbb{Z}_{\geq 0},\text{deg}\,P_n=n\}$
satisfy the three term recurrence relations,
$\eta P_n(\eta)=A_nP_{n+1}(\eta)+B_nP_n(\eta)+C_nP_{n-1}(\eta)$
($A_n,B_n,C_n$ : constants), and conversely the polynomials satisfying the
three term recurrence relations are orthogonal polynomials (Favard's
theorem \cite{szego}).
Since the exceptional orthogonal polynomials are not ordinary orthogonal
polynomials, they do not satisfy the three term recurrence relations.
In a previous paper \cite{rrmiop}, we showed that $M$-indexed orthogonal
polynomials $P_{\mathcal{D},n}(\eta)$ of Laguerre, Jacobi, Wilson and
Askey-Wilson types satisfy $3+2M$ term recurrence relations
($\mathcal{D}=\{d_1,\ldots,d_M\}$),
\begin{equation}
  R^{[M]}_{n,0}(\eta)P_{\mathcal{D},n}(\eta)=
  -\sum_{\genfrac{}{}{0pt}{}{k=-M-1}{k\neq 0}}^{M+1}
  R^{[M]}_{n,k}(\eta)P_{\mathcal{D},n+k}(\eta),
  \label{RRP}
\end{equation}
where $R^{[M]}_{n,k}(\eta)\,$'s are polynomials of degree $M+1-|k|$ in $\eta$.
In contrast to the three term recurrence relations, the coefficients of
\eqref{RRP} are not constants.
The three term recurrence relations are used to study bispectral properties
or dual polynomials \cite{szego,GH2}, in which the constant coefficients
of the recurrence relations are important.
To study bispectral properties, recurrence relations with constant
coefficients are desired,
\begin{equation}
 X(\eta)P_{\mathcal{D},n}(\eta)
 =\sum_{k=-L}^Lr_{n,k}^{X,\mathcal{D}}P_{\mathcal{D},n+k}(\eta)
 \ \ (\forall n\in\mathbb{Z}_{\geq 0}),
 \label{XP}
\end{equation}
where $r_{n,k}^{X,\mathcal{D}}\,$'s are constants and $X(\eta)$ is some
polynomial of degree $L$ in $\eta$.
Such recurrence relations for $M=1$ case were first given by Sasaki,
Tsujimoto and Zhedanov \cite{stz}. They found $1+4\ell$ term recurrence
relations. Recently Miki and Tsujimoto found different recurrence relations
with $3+2\ell$ terms \cite{mt}.
Their choices of $X(\eta)$ are $\Xi_{\ell}(\eta)^2$ and
$\int_0^{\eta}\Xi_{\ell}(y)dy$, respectively.
Dur\'{a}n studied recurrence relations with constant coefficients for
several exceptional orthogonal polynomials including exceptional Laguerre
polynomials by using duality \cite{duran}. 
The exceptional Laguerre polynomials in \cite{duran} correspond to eigenstates
and type $\I$ virtual states deletion, and our multi-indexed polynomials
correspond to type $\I$ and $\II$ virtual states deletion
(see \S\,\ref{sec:summary}).

In this paper we present infinitely many (conjectures of) recurrence relations
with constant coefficients for $M$-indexed orthogonal polynomials of Laguerre,
Jacobi, Wilson and Askey-Wilson types, namely we present infinitely many
polynomials $X(\eta)$ leading to \eqref{XP}.
The minimal degree of $X(\eta)$ is (conjectured as) $\ell_{\mathcal{D}}+1$,
where $\ell_{\mathcal{D}}$ is the degree of the lowest member multi-indexed
orthogonal polynomial $P_{\mathcal{D},0}(\eta)$, and this gives
$3+2\ell_{\mathcal{D}}$ term recurrence relations with constant coefficients.

This paper is organized as follows.
In section \ref{sec:method} we recapitulate some fundamental formulas of
the multi-indexed orthogonal polynomials and present a method deriving
recurrence relations with constant coefficients.
Section \ref{sec:RR} is the main part of the paper.
After discussing a necessary condition for $X(\eta)$, we present (conjectures
of) recurrence relations with constant coefficients for the multi-indexed
orthogonal polynomials of Laguerre, Jacobi, Wilson and Askey-Wilson types,
Conjecture\,\ref{conj_oQM} and Conjecture\,\ref{conj_idQM}.
The final section is for a summary and comments.
Some useful formulas of the multi-indexed orthogonal polynomials of
Laguerre, Jacobi, Wilson and Askey-Wilson types are listed in Appendix
\ref{sec:A}.
Some examples are presented in Appendix \ref{sec:B}.

\section{Method}
\label{sec:method}

In this section we explain an idea for deriving the recurrence relations with
constant coefficients.
We follow the notation of \cite{rrmiop}.
The virtual state wavefunction $\tilde{\phi}(x)$ is characterized by
the degree $\text{v}$ and the type $\text{t}$ ($\I$ or $\II$), like
$\tilde{\phi}_\text{v}^{\text{t}}(x)$.
For simplicity, we suppress type $\text{t}$ in many places.

The fundamental formulas of the multi-indexed orthogonal polynomials of
Laguerre, Jacobi, Wilson and Askey-Wilson types are found in \cite{rrmiop}.
Among them we recall that
\begin{align}
  &\hat{\mathcal{A}}_{d_1\ldots d_s}\phi_{d_1\ldots d_{s-1}\,n}(x)
  =\phi_{d_1\ldots d_s\,n}(x),\quad
  \hat{\mathcal{A}}_{d_1\ldots d_s}^{\dagger}\phi_{d_1\ldots d_s\,n}(x)
  =(\mathcal{E}_n-\tilde{\mathcal{E}}_{d_s})\phi_{d_1\ldots d_{s-1}\,n}(x),
  \label{AhDphiDn=}\\
  &\phi_{d_1\ldots d_s\,n}(x)
  =\Psi_{d_1\ldots d_s}(x)P_{d_1\ldots d_s,n}\big(\eta(x)\bigr)
  \ \ (n\in\mathbb{Z}_{\geq 0}),\quad
  P_{d_1\ldots d_s,n}(\eta)\eqdef 0\ \ (n<0),
  \label{phi=PsiP}\\
  &\text{deg}\,P_{d_1\dots d_s,n}(\eta)=\ell_{d_1\ldots d_s}+n,\quad
  \text{deg}\,\Xi_{d_1\dots d_s}(\eta)=\ell_{d_1\ldots d_s},\quad d_j>0,\n
  &\ \ \ell_{d_1\ldots d_s}=\sum_{j=1}^sd_j-\tfrac12s(s-1)+2s_{\I}s_{\II},
  \ \ s_{\text{t}}=\#\{d_j|\text{$d_j$:\,type $\text{t}$},j=1,\ldots,s\}
  \ (\text{t}=\I,\II),\\[-5pt]
  &(\phi_{d_1\ldots d_s\,n},\phi_{d_1\ldots d_s\,m})
  =(\Psi_{d_1\ldots d_s}^2P_{d_1\ldots d_s,n},P_{d_1\ldots d_s,m})
  =h_{d_1\ldots d_s,n}\delta_{nm},\n
  &\quad h_{d_1\ldots d_s,n}
  =\prod_{j=1}^s(\mathcal{E}_n-\tilde{\mathcal{E}}_{d_j})\cdot h_n.
  \label{hDn}
\end{align}
The relations \eqref{AhDphiDn=} are rewritten by using the step forward
($\hat{\mathcal{F}}$) and backward ($\hat{\mathcal{B}}$) shift operators as
\begin{equation}
  \hat{\mathcal{F}}_{d_1\ldots d_s}P_{d_1\ldots d_{s-1},n}(\eta)
  =P_{d_1\ldots d_s,n}(\eta),\quad
  \hat{\mathcal{B}}_{d_1\ldots d_s}P_{d_1\ldots d_s,n}(\eta)
  =(\mathcal{E}_n-\tilde{\mathcal{E}}_{d_s})P_{d_1\ldots d_{s-1},n}(\eta).
  \label{FDPDn=}
\end{equation}
Here $\hat{\mathcal{F}}_{d_1\ldots d_s}$ and
$\hat{\mathcal{B}}_{d_1\ldots d_s}$ are defined by
\begin{equation}
  \hat{\mathcal{F}}_{d_1\ldots d_s}\eqdef
  \Psi_{d_1\ldots d_s}(x)^{-1}\circ\hat{\mathcal{A}}_{d_1\ldots d_s}\circ
  \Psi_{d_1\ldots d_{s-1}}(x),
  \ \ \hat{\mathcal{B}}_{d_1\ldots d_s}\eqdef
  \Psi_{d_1\ldots d_{s-1}}(x)^{-1}\circ
  \hat{\mathcal{A}}_{d_1\ldots d_s}^{\dagger}\circ
  \Psi_{d_1\ldots d_s}(x),
  \label{FhD=}
\end{equation}
and their explicit forms are given in \eqref{FhD}--\eqref{BhD} and
\eqref{FhD_idQM}--\eqref{BhD_idQM}.
This gives Rodrigues type formula,
$P_{d_1\ldots d_s,n}(\eta)=\hat{\mathcal{F}}_{d_1\ldots d_s}\cdots
\hat{\mathcal{F}}_{d_1d_2}\hat{\mathcal{F}}_{d_1}P_n(\eta)$.
These formulas were not presented explicitly in our previous papers.
We remark that, from their explicit forms, $\hat{\mathcal{F}}_{d_1\ldots d_s}$
and $\hat{\mathcal{B}}_{d_1\ldots d_s}$ map rational functions of $\eta$
to rational functions of $\eta$.
For an appropriate parameter range (for example, see \cite{os25,os27,dp15}),
the Hamiltonians
$\mathcal{H}_{d_1\ldots d_s}$ are non-singular and their eigenfunctions
$\{\phi_{d_1\ldots d_s\,n}(x)\}_{n=0}^{\infty}$ form a complete set of the
Hilbert space.
For any polynomial $X(\eta)$ in $\eta$, the function
$X\bigl(\eta(x)\bigr)\phi_{d_1\dots d_s\,n}(x)$ belongs to the Hilbert space.

The Hamiltonian $\mathcal{H}_{d_1\ldots d_s}$ does not depend on the order of
$d_j$'s.
On the other hand, the multi-indexed orthogonal polynomial
$P_{d_1\ldots d_s,n}(\eta)$ changes the sign under a permutation of $d_j$'s,
$P_{d_{\sigma_1}\ldots d_{\sigma_s},n}(\eta)
=\text{sgn}\genfrac{(}{)}{0pt}{}{1\ \ldots\ s\ }{\sigma_1\,\ldots\,\sigma_s}
P_{d_1\ldots d_s,n}(\eta)$.
The denominator polynomial $\Xi_{d_1\ldots d_s}(\eta)$ also changes the sign,
$\Xi_{d_{\sigma_1}\ldots d_{\sigma_s}}(\eta)
=\text{sgn}\genfrac{(}{)}{0pt}{}{1\ \ldots\ s\ }{\sigma_1\,\ldots\,\sigma_s}
\Xi_{d_1\ldots d_s}(\eta)$.
We write $\mathcal{H}_{d_1\ldots d_M}$, $\phi_{d_1\ldots d_M\,n}(x)$,
$P_{d_1\ldots d_M,n}(\eta)$, $\Xi_{d_1\ldots d_M}(\eta)$,
$h_{d_1\dots d_M,n}$, etc.$\!$\, as
$\mathcal{H}_{\mathcal{D}}$, $\phi_{\mathcal{D}\,n}(x)$,
$P_{\mathcal{D},n}(\eta)$, $\Xi_{\mathcal{D}}(\eta)$, $h_{\mathcal{D},n}$,
etc., respectively ($\mathcal{D}=\{d_1,\ldots,d_M\}$).\footnote{
In the Appendix of \cite{rrmiop}, we assumed the `standard order'
$\mathcal{D}=\{d^{\I}_1,\ldots,d^{\I}_{M_{\I}},
d^{\II}_1,\ldots,d^{\II}_{M_{\II}}\}$ for simplicity. We do not assume it
in this paper.
}

First we note the following property of orthogonal polynomials.
\begin{lemma}
Let us assume for a certain polynomial $X(\eta)$ of degree $L$ in $\eta$ that
\begin{equation}
  X(\eta)P_{\mathcal{D},n}(\eta)=\sum_{k=-n}^Lr_{n,k}^{X,\mathcal{D}}
  P_{\mathcal{D},n+k}(\eta)\ \ (\forall n\in\mathbb{Z}_{\geq 0}).
  \label{XPDn}
\end{equation}
Here $r_{n,k}^{X,\mathcal{D}}$'s are constants.
The sum $\sum\limits_{k=-n}^L$ can be replaced by $\sum\limits_{k=-L}^L$.
\label{lemma1}
\end{lemma}
\underline{Proof}
Multiplying by $\Psi_{\mathcal{D}}(x)$ to \eqref{XPDn}, we have
\begin{equation*}
  X(\eta)\phi_{\mathcal{D},n}(x)
  =\sum_{k=-n}^Lr_{n,k}^{X,\mathcal{D}}\phi_{\mathcal{D}\,n+k}(x).
\end{equation*}
By using \eqref{hDn} we have
\begin{align}
  &\quad(\phi_{\mathcal{D}\,m},X\phi_{\mathcal{D}\,n})
  =\sum_{k=-n}^Lr_{n,k}^{X,\mathcal{D}}h_{\mathcal{D},m}\delta_{m,n+k}
  =\theta(m\leq n+L)\,r_{n,m-n}^{X,\mathcal{D}}h_{\mathcal{D},m}\n
  &=(X\phi_{\mathcal{D}\,m},\phi_{\mathcal{D}\,n})
  =\sum_{k=-m}^Lr_{m,k}^{X,\mathcal{D}}h_{\mathcal{D},n}\delta_{n,m+k}
  =\theta(m\geq n-L)\,r_{m,n-m}^{X,\mathcal{D}}h_{\mathcal{D},n},
\end{align}
where $\theta(P)$ is a step function for a proposition $P$,
$\theta(P)=1$ ($P$ : true), $0$ ($P$ : false).
This means $(\phi_{\mathcal{D}\,m},X\phi_{\mathcal{D}\,n})=0$
unless $n-L\leq m\leq n+L$.
Namely $r_{n,k}^{X,\mathcal{D}}=0$ unless $-L\leq k\leq L$.
\hfill\fbox{}\medskip

\noindent
{\bf Remark}$\,$
Although the inner product formulas used in the proof are valid only for `real'
$X(\eta)$ ($X^*=X$) and an appropriate parameter range such that the Hamiltonian
is non-singular, the final result, which represents the polynomial equations,
is valid for any parameter values and complex $X(\eta)$.
\medskip

Next we explain a method to obtain recurrence relations with constant
coefficients.
Let $X(\eta)$ be a polynomial of degree $L$ in $\eta$.
Since $X(\eta)\phi_{\mathcal{D}\,n}(x)$ belongs to the Hilbert space and
$\{\phi_{\mathcal{D}\,n}(x)\}_{n=0}^{\infty}$ is a complete set,
we have the expansion
\begin{equation}
  X(\eta)\phi_{\mathcal{D}\,n}(x)
  =\sum_{k=-n}^{\infty}r_{n,k}^{X,\mathcal{D}}\phi_{\mathcal{D}\,n+k}(x),
  \label{XphiDn=}
\end{equation}
where $r_{n,k}^{X,\mathcal{D}}\,$'s are constants.
By using this and \eqref{hDn}, we obtain
\begin{equation}
  (\phi_{\mathcal{D}\,m},X\phi_{\mathcal{D}\,n})
  =\sum_{k=-n}^{\infty}r_{n,k}^{X,\mathcal{D}}h_{\mathcal{D},m}\delta_{m,n+k}
  =r_{n,m-n}^{X,\mathcal{D}}h_{\mathcal{D},m}.
  \label{(phiDm,XphiDn)}
\end{equation}
On the other hand, by using \eqref{AhDphiDn=}--\eqref{phi=PsiP} and
\eqref{FhD=}, we obtain
\begin{align}
  &\quad(\phi_{\mathcal{D}\,m},X\phi_{\mathcal{D}\,n})\n
  &=(\hat{\mathcal{A}}_{d_1\ldots d_M}\cdots\hat{\mathcal{A}}_{d_1d_2}
  \hat{\mathcal{A}}_{d_1}\phi_m,
  X\phi_{\mathcal{D}\,n})\n
  &=\bigl(\phi_m,\hat{\mathcal{A}}_{d_1}^{\dagger}
  \hat{\mathcal{A}}_{d_1d_2}^{\dagger}\cdots
  \hat{\mathcal{A}}_{d_1\ldots d_M}^{\dagger}
  (X\phi_{\mathcal{D}\,n})\bigr)\n
  &=\bigl(\phi_0P_m,(\phi_0\hat{\mathcal{B}}_{d_1}\Psi_{d_1}^{-1})
  (\Psi_{d_1}\hat{\mathcal{B}}_{d_1d_2}\Psi_{d_1d_2}^{-1})\cdots
  (\Psi_{d_1\ldots d_{M-1}}\hat{\mathcal{B}}_{d_1\ldots d_M}
  \Psi_{d_1\ldots d_M}^{-1})
  (\Psi_{d_1\ldots d_M}XP_{\mathcal{D},n})\bigr)\n
  &=\bigl(\phi_0P_m,\phi_0\hat{\mathcal{B}}_{d_1}
  \hat{\mathcal{B}}_{d_1d_2}\cdots
  \hat{\mathcal{B}}_{d_1\ldots d_M}
  (XP_{\mathcal{D},n})\bigr)\n
  &=\bigl(\phi_0^2P_m,\hat{\mathcal{B}}_{d_1}
  \hat{\mathcal{B}}_{d_1d_2}\cdots
  \hat{\mathcal{B}}_{d_1\ldots d_M}
  (XP_{\mathcal{D},n})\bigr).
  \label{(phiDm,XphiDn)2}
\end{align}
{}From the property of $\hat{\mathcal{B}}_{d_1\ldots d_s}$, the function
$\hat{\mathcal{B}}_{d_1}\hat{\mathcal{B}}_{d_1d_2}\cdots
\hat{\mathcal{B}}_{d_1\ldots d_M}(XP_{\mathcal{D},n})$ is a rational function
of $\eta$.
When it is not a polynomial in $\eta$, we have infinitely many $m$ such that
$(\phi_{\mathcal{D}\,m},X\phi_{\mathcal{D}\,n})\neq 0$, namely r.h.s.$\!$\, of
\eqref{XphiDn=} is an infinite sum.
Let us consider the case that
$\hat{\mathcal{B}}_{d_1}\hat{\mathcal{B}}_{d_1d_2}\cdots
\hat{\mathcal{B}}_{d_1\ldots d_M}$ $(XP_{\mathcal{D},n})$
is a polynomial of degree $n+L'$ in $\eta$.
Since any polynomial in $\eta$ can be expanded in $P_n(\eta)$, we have
\begin{equation}
  \hat{\mathcal{B}}_{d_1}\hat{\mathcal{B}}_{d_1d_2}\cdots
  \hat{\mathcal{B}}_{d_1\ldots d_M}(XP_{\mathcal{D},n})
  =\sum_{k=-n}^{L'}r_{n,k}^{(0)\,X,\mathcal{D}}P_{n+k}(\eta),
  \label{BBBXP}
\end{equation}
where $r_{n,k}^{(0)\,X,\mathcal{D}}\,$'s are constants.
Substituting this to \eqref{(phiDm,XphiDn)2}, we obtain
\begin{equation}
  (\phi_{\mathcal{D}\,m},X\phi_{\mathcal{D}\,n})
  =\sum_{k=-n}^{L'}r_{n,k}^{(0)\,X,\mathcal{D}}h_m\delta_{m\,n+k},
  =\theta(m\leq L'+n)\,r_{n,m-n}^{(0)X,\mathcal{D}}h_m.
  \label{(phiDm,XphiDn)3}
\end{equation}
Eqs.\eqref{(phiDm,XphiDn)} and \eqref{(phiDm,XphiDn)3} imply
\begin{equation}
  r_{n,k}^{X,\mathcal{D}}=0\ \ (k>L'),\quad
  r_{n,k}^{X,\mathcal{D}}h_{\mathcal{D},n+k}
  =r_{n,k}^{(0)\,X,\mathcal{D}}h_{n+k}\ \ (-n\leq k\leq L').
\end{equation}
Thus we obtain
\begin{equation}
  X(\eta)\phi_{\mathcal{D}\,n}(x)
  =\sum_{k=-n}^{L'}r_{n,k}^{X,\mathcal{D}}\phi_{\mathcal{D}\,n+k}(x),
\end{equation}
namely,
\begin{equation}
  X(\eta)P_{\mathcal{D},n}(\eta)
  =\sum_{k=-n}^{L'}r_{n,k}^{X,\mathcal{D}}P_{\mathcal{D},n+k}(\eta).
\end{equation}
By comparing the degree of both sides, we have $L'=L$.
By Lemma\,\ref{lemma1}, the sum $\sum\limits_{k=-n}^L$ can be replaced
by $\sum\limits_{k=-L}^L$.

We summarize this argument as the following proposition.
\begin{prop}
Let us assume for a certain polynomial $X(\eta)$ of degree $L$ in $\eta$
that the function 
$\hat{\mathcal{B}}_{d_1}\hat{\mathcal{B}}_{d_1d_2}\cdots
\hat{\mathcal{B}}_{d_1\ldots d_M}(XP_{\mathcal{D},n})$
is a polynomial in $\eta$. Expand it as \eqref{BBBXP}.
We have $1+2L$ term recurrence relations with constant coefficients
for $P_{\mathcal{D},n}(\eta)$ :
\begin{equation}
  X(\eta)P_{\mathcal{D},n}(\eta)
  =\sum_{k=-L}^{L}r_{n,k}^{X,\mathcal{D}}P_{\mathcal{D},n+k}(\eta)
  \ \ (\forall n\in\mathbb{Z}_{\geq 0}),\quad
  r_{n,k}^{X,\mathcal{D}}=\frac{r_{n,k}^{(0)\,X,\mathcal{D}}}
  {\prod_{j=1}^M(\mathcal{E}_{n+k}-\tilde{\mathcal{E}}_{d_j})}.
  \label{RRX}
\end{equation}
\label{rrXPDn}
\end{prop}
\vspace*{-4mm}
{\bf Remark 1}$\,$
See Remark below Lemma\,\ref{lemma1}.\\
{\bf Remark 2}$\,$
Under the assumption of this proposition, the functions
$\hat{\mathcal{B}}_{d_1\ldots d_j}\hat{\mathcal{B}}_{d_1\ldots d_jd_{j+1}}
\cdots\hat{\mathcal{B}}_{d_1\ldots d_M}$ $(XP_{\mathcal{D},n})$
$(j=2,\ldots,M)$ are also polynomials in $\eta$.\\
{\bf Remark 3}$\,$
The function $\hat{\mathcal{B}}_{d_1}\hat{\mathcal{B}}_{d_1d_2}\cdots
\hat{\mathcal{B}}_{d_1\ldots d_M}(XP_{\mathcal{D},n})$ is rewritten as
\begin{equation}
  \hat{\mathcal{B}}_{d_1}\hat{\mathcal{B}}_{d_1d_2}\cdots
  \hat{\mathcal{B}}_{d_1\ldots d_M}(XP_{\mathcal{D},n})
  =(\hat{\mathcal{B}}_{d_1}\hat{\mathcal{B}}_{d_1d_2}\cdots
  \hat{\mathcal{B}}_{d_1\ldots d_M}X
  \hat{\mathcal{F}}_{d_1\ldots d_M}\cdots\hat{\mathcal{F}}_{d_1d_2}
  \hat{\mathcal{F}}_{d_1})P_n.
\end{equation}
This operator $\hat{\mathcal{B}}_{d_1}\cdots X\cdots\hat{\mathcal{F}}_{d_1}$
maps polynomials in $\eta$ to rational functions of $\eta$.
To find a proper polynomial $X(\eta)$ giving recurrence relations with constant
coefficients is rephrased as follows;
Find a polynomial $X(\eta)$ such that the operator
$\hat{\mathcal{B}}_{d_1}\hat{\mathcal{B}}_{d_1d_2}\cdots
\hat{\mathcal{B}}_{d_1\ldots d_M}X\hat{\mathcal{F}}_{d_1\ldots d_M}
$ $\cdots\hat{\mathcal{F}}_{d_1d_2}\hat{\mathcal{F}}_{d_1}$
maps polynomials in $\eta$ to polynomials in $\eta$.

\section{Recurrence Relations with Constant Coefficients}
\label{sec:RR}

In this section we present recurrence relations with constant coefficients
\eqref{XP} for the multi-indexed orthogonal polynomials of Laguerre, Jacobi,
Wilson and Askey-Wilson types.

\subsection{Multi-indexed Laguerre and Jacobi polynomials}
\label{sec:RRoQM}

In this subsection we discuss the recurrence relations with constant
coefficients for the multi-indexed Laguerre and Jacobi polynomials.
We note that the first order differential operator of the form
$a(x)\frac{d}{dx}+b(x)$
($a(x),b(x)$ : functions of $x$) acts on the product of two functions
$f(x)$ and $g(z)$ as
\begin{equation}
  \Bigl(a(x)\frac{d}{dx}+b(x)\Bigr)\bigl(f(x)g(x)\bigr)
  =f(x)\Bigl(a(x)\frac{d}{dx}+b(x)\Bigr)g(x)+a(x)\frac{df(x)}{dx}g(x).
  \label{act_fg}
\end{equation}

First we consider a necessary condition for $X(\eta)$ giving recurrence
relations with constant coefficients.
Let us assume \eqref{XPDn} for a polynomial $X(\eta)$ of degree $L$ in $\eta$.
Applying $\hat{\mathcal{B}}_{\mathcal{D}}=\hat{\mathcal{B}}_{d_1\ldots d_M}$
\eqref{BhD} to \eqref{XPDn}, we have
\begin{align}
  &\quad\hat{\mathcal{B}}_{\mathcal{D}}
  \bigl(X(\eta)P_{\mathcal{D},n}(\eta)\bigr)
  =\sum_{k=-n}^Lr_{n,k}^{X,\mathcal{D}}\hat{\mathcal{B}}_{\mathcal{D}}
  P_{\mathcal{D},n+k}(\eta)
  =\sum_{k=-n}^Lr_{n,k}^{X,\mathcal{D}}
  (\mathcal{E}_{n+k}-\tilde{\mathcal{E}}_{d_M})
  P_{d_1\ldots d_{M-1},n}(\eta)\n
  &=X(\eta)\hat{\mathcal{B}}_{\mathcal{D}}P_{\mathcal{D},n}(\eta)
  -\cF^2e^{\hat{\mathcal{B}}}_{\mathcal{D}}(\eta)
  \frac{\Xi_{d_1\ldots d_{M-1}}(\eta)}{\Xi_{\mathcal{D}}(\eta)}
  \frac{dX(\eta)}{d\eta}P_{\mathcal{D},n}(\eta)\n
  &=(\mathcal{E}_n-\tilde{\mathcal{E}}_{d_M})X(\eta)
  P_{d_1\ldots d_{M-1},n}(\eta)
  -\cF^2e^{\hat{\mathcal{B}}}_{\mathcal{D}}(\eta)
  \frac{\Xi_{d_1\ldots d_{M-1}}(\eta)}{\Xi_{\mathcal{D}}(\eta)}
  \frac{dX(\eta)}{d\eta}P_{\mathcal{D},n}(\eta),
  \label{BhDXPDn}
\end{align}
where \eqref{act_fg} and \eqref{FDPDn=} are used.
Since the expression in the first line is a polynomial in $\eta$,
the expression in the last line should be so.
The denominator polynomial $\Xi_{\mathcal{D}}(\eta)$ does not have common roots
with $e^{\hat{\mathcal{B}}}_{\mathcal{D}}(\eta)$ and
$P_{\mathcal{D},n}(\eta)$ for some $n$.
Therefore, if $\Xi_{\mathcal{D}}(\eta)=\Xi_{d_1\ldots d_M}(\eta)$ does not
have common roots with $\Xi_{d_1\ldots d_{M-1}}(\eta)$, the polynomial
$\frac{dX(\eta)}{d\eta}$ should be divisible by $\Xi_{\mathcal{D}}(\eta)$.

We summarize this argument as follows.
\begin{prop}
Let $X(\eta)$ be a polynomial of degree $L$ in $\eta$.
Assume \eqref{XPDn} and
\begin{equation}
  \frac{dX(\eta)}{d\eta}
  =\Xi_{\mathcal{D}}(\eta)Y(\eta),\quad
  \text{\rm $Y(\eta)$ : a polynomial in $\eta$}.
  \label{Xcond}
\end{equation}
Then one action of $\hat{\mathcal{B}}_{\mathcal{D}}$ to the both sides of
\eqref{XPDn}
keeps the polynomiality intact.
\label{propXweak}
\end{prop}
{\bf Remark}$\,$
If two polynomials in $\eta$,
$\Xi_{\mathcal{D}}(\eta)=\Xi_{d_1\ldots d_M}(\eta)$ and
$\Xi_{d_1\ldots d_{M-1}}(\eta)$, do not have common roots,
the polynomial of degree $L$ in $\eta$, $X(\eta)$, satisfying \eqref{XPDn}
should satisfy \eqref{Xcond} for some polynomial $Y(\eta)$.

The overall normalization and the constant term of $X(\eta)$ are not important,
because the change of the former induces that of the overall normalization
of $r_{n,k}^{X,\mathcal{D}}$ and the shift of the latter induces that of
$r_{n,0}^{X,\mathcal{D}}$.
By taking the constant term of $X(\eta)$ as $X(0)=0$, the condition for the
candidate of $X(\eta)$ \eqref{Xcond} gives
\begin{equation}
  X(\eta)=\int_0^{\eta}\Xi_{\mathcal{D}}(y)Y(y)dy,\quad
  \text{deg}\,X(\eta)=L=\ell_{\mathcal{D}}+\text{deg}\,Y(\eta)+1.
  \label{X=int}
\end{equation}
The minimal degree candidate of $X(\eta)$, which corresponds to $Y(\eta)=1$, is
\begin{equation}
  X_{\text{min}}(\eta)=\int_0^{\eta}\Xi_{\mathcal{D}}(y)dy,\quad
  \text{deg}\,X_{\text{min}}(\eta)=\ell_{\mathcal{D}}+1.
  \label{Xmin}
\end{equation}

Based on these properties we present our main result.
After one action of $\hat{\mathcal{B}}_{\mathcal{D}}$ to \eqref{XPDn},
further actions of
$\hat{\mathcal{B}}_{d_1\ldots d_{M-1}},\hat{\mathcal{B}}_{d_1\ldots d_{M-2}},
\ldots$ give more conditions for $X(\eta)$. However, it seems that these
additional conditions are satisfied automatically by the original condition
\eqref{Xcond} and by the properties of $P_{d_1\ldots d_s,n}(\eta)$ and
$\Xi_{d_1\ldots d_s}(\eta)$,
for example, see the proof of $M=2$ case in Remark 5 below.
We conjecture that this candidate $X(\eta)$ \eqref{X=int} actually gives
recurrence relations with constant coefficients.
\begin{conj}
For any polynomial $Y(\eta)$, we take $X(\eta)$ as \eqref{X=int}.
Then the multi-indexed Laguerre and Jacobi polynomials $P_{\mathcal{D},n}(\eta)$
satisfy $1+2L$ term recurrence relations with constant coefficients \eqref{XP}.
\label{conj_oQM}
\end{conj}
{\bf Remark 1}$\,$
If two polynomials in $\eta$,
$\Xi_{\mathcal{D}}(\eta)=\Xi_{d_1\ldots d_M}(\eta)$ and
$\Xi_{d_1\ldots d_{M-1}}(\eta)$, do not have common roots,
this conjecture exhausts all possible
$X(\eta)$ giving recurrence relations with constant coefficients, and
the minimal degree choice $X(\eta)=X_{\text{min}}(\eta)$ \eqref{Xmin} gives
$3+2\ell_{\mathcal{D}}$ term recurrence relations.\\
{\bf Remark 2}$\,$
The minimal degree polynomial $X_{\text{min}}(\eta)$ can be divisible by $\eta$.
The degree of $\frac{X_{\text{min}}(\eta)}{\eta}$ is $\ell_{\mathcal{D}}$,
which is the lowest degree of $P_{\mathcal{D},n}(\eta)$.
The recurrence relations \eqref{XP},
$\frac{X_{\text{min}}(\eta)}{\eta}\times\eta P_{\mathcal{D},n}(\eta)
=\sum\cdots$,
can be regarded as a natural generalization of the three term recurrence
relations of the ordinary orthogonal polynomial $P_n(\eta)$,
$1\times\eta P_n(\eta)=A_nP_{n+1}(\eta)+B_nP_n(\eta)+C_nP_{n-1}(\eta)$.\\
{\bf Remark 3}$\,$
Since $Y(\eta)$ is arbitrary, we obtain infinitely many recurrence relations.
However not all of them are independent.
For `$M=0$ case' (namely, ordinary orthogonal polynomials), it is trivial that
recurrence relations obtained from arbitrary $Y(\eta)$ ($\text{deg}\,Y\geq 1$)
are derived by the three term recurrence relations.\\
{\bf Remark 4}$\,$
For $M=1$ case ($\mathcal{D}=\{\ell\}$), the minimal degree choice
$X(\eta)=X_{\text{min}}(\eta)$, which gives $3+2\ell$ term recurrence
relations, was given by Miki and Tsujimoto \cite{mt}, and the choice
$X(\eta)=\Xi_{\ell}(\eta)^2$, which corresponds to $Y(\eta)=2\Xi_{\ell}(\eta)$
and gives $1+4\ell$ term recurrence relations, was given by Sasaki,
Tsujimoto and Zhedanov \cite{stz}.
For general $M$, $Y(\eta)=2\partial_{\eta}\Xi_{\mathcal{D}}(\eta)p(\eta)
+\Xi_{\mathcal{D}}(\eta)\partial_{\eta}p(\eta)$, where $p(\eta)$ is any
polynomial in $\eta$, gives $X(\eta)=\Xi_{\mathcal{D}}(\eta)^2p(\eta)$.\\
{\bf Remark 5}$\,$
Direct verification of this conjecture is rather straightforward for lower $M$
and smaller $d_j$, $n$ and $\text{deg}\,Y$, by a computer algebra system,
e.g.\! Mathematica.
The coefficients $r_{n,k}^{X,\mathcal{D}}$ are explicitly obtained for
small $d_j$ and $n$. However, to obtain the closed expression of
$r_{n,k}^{X,\mathcal{D}}$ for general $n$ is not an easy task even for
small $d_j$, and it is a different kind of problem. 
We present some examples in Appendix \ref{sec:B}.\\
{\bf Remark 6}$\,$
For $M=1,2$ we can prove this conjecture.
Since we have Proposition\,\ref{rrXPDn}, it is sufficient to show that
$\hat{\mathcal{B}}_{d_1}\hat{\mathcal{B}}_{d_1d_2}\cdots
\hat{\mathcal{B}}_{d_1\ldots d_M}(XP_{\mathcal{D},n})$
is a polynomial in $\eta$.\\
\underline{$M=1$}
{}From \eqref{BhDXPDn} we have
\begin{equation*}
  \hat{\mathcal{B}}_{d_1}\bigl(X(\eta)P_{d_1,n}(\eta)\bigr)
  =(\mathcal{E}_n-\tilde{\mathcal{E}}_{d_1})X(\eta)P_n(\eta)
  -\cF^2e^{\hat{\mathcal{B}}}_{d_1}(\eta)Y(\eta)P_{d_1,n}(\eta).
\end{equation*}
This is a polynomial in $\eta$. Thus $M=1$ case is proved.
\hfill\fbox{}\\
\underline{$M=2$} {}From \eqref{BhDXPDn} and \eqref{BhD} we have
\begin{align*}
  &\quad\hat{\mathcal{B}}_{d_1}\hat{\mathcal{B}}_{d_1d_2}
  \bigl(X(\eta)P_{d_1d_2,n}(\eta)\bigr)\n
  &=\hat{\mathcal{B}}_{d_1}\bigl(
  (\mathcal{E}_n-\tilde{\mathcal{E}}_{d_2})X(\eta)P_{d_1,n}(\eta)
  -\cF^2e^{\hat{\mathcal{B}}}_{d_1d_2}(\eta)\Xi_{d_1}(\eta)
  Y(\eta)P_{d_1d_2,n}(\eta)\bigr)\n
  &=(\mathcal{E}_n-\tilde{\mathcal{E}}_{d_2})\Bigl(
  (\mathcal{E}_n-\tilde{\mathcal{E}}_{d_1})X(\eta)P_n(\eta)
  -\cF^2\frac{e^{\hat{\mathcal{B}}}_{d_1}(\eta)}
  {\Xi_{d_1}(\eta)}\Xi_{d_1d_2}(\eta)
  Y(\eta)P_{d_1,n}(\eta)\Bigr)\n
  &\quad+\cF^4e^{\hat{\mathcal{B}}}_{d_1}(\eta)
  \partial_{\eta}\bigl(e^{\hat{\mathcal{B}}}_{d_1d_2}(\eta)
  Y(\eta)P_{d_1d_2,n}(\eta)\bigr)
  +\cF^4\frac{e^{\hat{\mathcal{B}}}_{d_1}(\eta)}{\Xi_{d_1}(\eta)}
  e^{\hat{\mathcal{B}}}_{d_1d_2}(\eta)
  \partial_{\eta}\Xi_{d_1}(\eta)Y(\eta)P_{d_1d_2,n}(\eta)\n
  &\quad-\cF^3\tilde{e}^{\hat{\mathcal{B}}}_{d_1}
  e^{\hat{\mathcal{B}}}_{d_1d_2}(\eta)Y(\eta)P_{d_1d_2,n}(\eta).
\end{align*}
By using \eqref{PDnB} with $s=2$, this becomes
\begin{align*}
  &=(\mathcal{E}_n-\tilde{\mathcal{E}}_{d_2})
  (\mathcal{E}_n-\tilde{\mathcal{E}}_{d_1})X(\eta)P_n(\eta)\n
  &\quad+\cF^4e^{\hat{\mathcal{B}}}_{d_1}(\eta)
  \partial_{\eta}\bigl(e^{\hat{\mathcal{B}}}_{d_1d_2}(\eta)
  Y(\eta)P_{d_1d_2,n}(\eta)\bigr)
  +\cF^4e^{\hat{\mathcal{B}}}_{d_1}(\eta)e^{\hat{\mathcal{B}}}_{d_1d_2}(\eta)
  Y(\eta)\partial_{\eta}P_{d_1d_2,n}(\eta)\n
  &\quad
  -\cF^3\bigl(\tilde{e}^{\hat{\mathcal{B}}}_{d_1}
  e^{\hat{\mathcal{B}}}_{d_1d_2}(\eta)
  +e^{\hat{\mathcal{B}}}_{d_1}(\eta)
  \tilde{e}^{\hat{\mathcal{B}}}_{d_1d_2}\bigr)
  Y(\eta)P_{d_1d_2,n}(\eta).
\end{align*}
This is a polynomial in $\eta$. Thus $M=2$ case is proved.
\hfill\fbox{}

\subsection{Multi-indexed Wilson and Askey-Wilson polynomials}
\label{sec:RRidQM}

In this subsection we discuss the recurrence relations with constant
coefficients for the multi-indexed Wilson and Askey-Wilson polynomials.
We restrict the parameters: $\{a_1^*,a_2^*\}=\{a_1,a_2\}$ (as a set) and
$\{a_3^*,a_4^*\}=\{a_3,a_4\}$ (as a set).

The sinusoidal coordinate $\eta(x)$ is $\eta(x)=x^2$ for Wilson case
and $\eta(x)=\cos x$ for Askey-Wilson case.
They satisfy \cite{os14}
\begin{equation}
  \frac{\eta(x-i\frac{\gamma}{2})^{n+1}-\eta(x+i\frac{\gamma}{2})^{n+1}}
  {\eta(x-i\frac{\gamma}{2})-\eta(x+i\frac{\gamma}{2})}
  =\sum_{k=0}^ng_n^{\prime\,(k)}\eta(x)^{n-k}
  \ \ (n\in\mathbb{Z}_{\geq 0}),
\end{equation}
where $g_n^{\prime\,(k)}$ is given by \cite{os32}
\begin{align}
  &\eta(x)=x^2:
  \ \ g_n^{\prime\,(k)}
  =\frac{(-1)^k}{2^{2k+1}}\genfrac{(}{)}{0pt}{}{2n+2}{2k+1},\n
  &\eta(x)=\cos x:
  \ \ g_n^{\prime\,(k)}=\theta(\text{$k$ : even})\frac{(n+1)!}{2^k}
  \sum_{r=0}^{\frac{k}{2}}\genfrac{(}{)}{0pt}{}{n-k+r}{r}
  \frac{(-1)^r[\![n-k+1+2r]\!]'}{(\frac{k}{2}-r)!\,(n-\frac{k}{2}+1+r)!},\n
  &\qquad\qquad\qquad\qquad\qquad
  [\![n]\!]'\eqdef\frac{e^{-\frac{\gamma}{2}n}-e^{\frac{\gamma}{2}n}}
  {e^{-\frac{\gamma}{2}}-e^{\frac{\gamma}{2}}}.
\end{align}
For a polynomial $p(\eta)$ in $\eta$, when it is regarded as a function of $x$,
we denote it by adding a check,
\begin{equation}
  \check{p}(x)\eqdef p\bigl(\eta(x)\bigr).  
\end{equation}
Since $\eta(x-i\frac{m}{2}\gamma)+\eta(x+i\frac{m}{2}\gamma)$ and
$\eta(x-i\frac{m}{2}\gamma)\eta(x+i\frac{m}{2}\gamma)$ ($m\in\mathbb{Z}$) are
expressed as polynomials in $\eta(x)$, any symmetric polynomial in
$\eta(x-i\frac{m}{2}\gamma)$ and $\eta(x+i\frac{m}{2}\gamma)$ is expressed
as a polynomial in $\eta(x)$ \cite{os14,rrmiop}.
For example, the followings are polynomials in $\eta(x)$ ($p,p_1,p_2$ :
polynomials in $\eta$):
\begin{align}
  &\check{p}(x-i\tfrac{\gamma}{2})+\check{p}(x+i\tfrac{\gamma}{2}),
  \ \ \frac{\check{p}(x-i\frac{\gamma}{2})-\check{p}(x+i\frac{\gamma}{2})}
  {\eta(x-i\frac{\gamma}{2})-\eta(x+i\frac{\gamma}{2})},\n
  &\check{p}(x-i\tfrac{\gamma}{2})\check{p}(x+i\tfrac{\gamma}{2}),
  \ \ \frac{\check{p}_1(x-i\gamma)\check{p}_2(x-i\frac{\gamma}{2})
  -\check{p}_1(x+i\gamma)\check{p}_2(x+i\frac{\gamma}{2})}
  {\eta(x-i\frac{\gamma}{2})-\eta(x+i\frac{\gamma}{2})},\n
 &\eta(x)=x^2:
  \ \ x\check{p}_1(x+i\tfrac{\gamma}{2})\check{p}_2(x-i\tfrac{\gamma}{2})
  +x\check{p}_1(x-i\tfrac{\gamma}{2})\check{p}_2(x+i\tfrac{\gamma}{2}),\n
  &\eta(x)=\cos x:
  \ \ e^{\pm ix}\check{p}_1(x+i\tfrac{\gamma}{2})
  \check{p}_2(x-i\tfrac{\gamma}{2})
  +e^{\mp ix}\check{p}_1(x-i\tfrac{\gamma}{2})
  \check{p}_2(x+i\tfrac{\gamma}{2}).
  \label{poly}
\end{align}
For a polynomial $p(\eta)$ in $\eta$, let us define a polynomial in $\eta$,
$I[p](\eta)$, as follows:
\begin{equation}
  p(\eta)=\sum_{k=0}^na_k\eta^k\mapsto
  I[p](\eta)=\sum_{k=0}^{n+1}b_k\eta^k,
\end{equation}
where $b_k$'s are defined by
\begin{equation}
  b_{k+1}=\frac{1}{g_k^{\prime\,(0)}}
  \Bigl(a_k-\sum_{j=k+1}^ng_j^{\prime\,(j-k)}b_{j+1}\Bigr)
  \ \ (k=n,n-1,\ldots,1,0),\quad
  b_0=0.
\end{equation}
The constant term of $I[p](\eta)$ is chosen to be zero.
It is easy to show that this polynomial $I[p](\eta)=P(\eta)$ satisfies
\begin{equation}
  \frac{\check{P}(x-i\frac{\gamma}{2})-\check{P}(x+i\frac{\gamma}{2})}
  {\eta(x-i\frac{\gamma}{2})-\eta(x+i\frac{\gamma}{2})}
  =\check{p}(x).
\end{equation}

The operator of the form $a(x)e^{\frac{\gamma}{2}p}-b(x)e^{-\frac{\gamma}{2}p}$
($a(x),b(x)$ : functions of $x$) acts on the product of two functions
$f(x)$ and $g(z)$ as
\begin{align}
  &\quad\bigl(a(x)e^{\frac{\gamma}{2}p}-b(x)e^{-\frac{\gamma}{2}p}\bigr)
  \bigl(f(x)g(x)\bigr)\n
  &=a(x)f(x-i\tfrac{\gamma}{2})g(x-i\tfrac{\gamma}{2})
  -b(x)f(x+i\tfrac{\gamma}{2})g(x+i\tfrac{\gamma}{2})\n
  &=f^{(+)}(x)\bigl(a(x)g(x-i\tfrac{\gamma}{2})
  -b(x)g(x+i\tfrac{\gamma}{2})\bigr)
  -if^{(-)}(x)\bigl(a(x)g(x-i\tfrac{\gamma}{2})
  +b(x)g(x+i\tfrac{\gamma}{2})\bigr)\n
  &=f^{(+)}(x)\bigl(a(x)e^{\frac{\gamma}{2}p}
  -b(x)e^{-\frac{\gamma}{2}p}\bigr)g(x)
  -if^{(-)}(x)\bigl(a(x)g(x-i\tfrac{\gamma}{2})
  +b(x)g(x+i\tfrac{\gamma}{2})\bigr),
  \label{act_fg_idQM}
\end{align}
where $f^{(\pm)}(x)$ are defined by \cite{rrmiop}
\begin{equation}
  f^{(+)}(x)\eqdef\frac12\bigl(f(x-i\tfrac{\gamma}{2})
  +f(x+i\tfrac{\gamma}{2})\bigr),\quad
  f^{(-)}(x)\eqdef\frac{i}{2}\bigl(f(x-i\tfrac{\gamma}{2})
  -f(x+i\tfrac{\gamma}{2})\bigr).
\end{equation}
The auxiliary function $\varphi(x)$ is rewritten as
\begin{equation}
  \varphi(x)=ic_{\varphi}\bigl(\eta(x-i\tfrac{\gamma}{2})
  -\eta(x-i\tfrac{\gamma}{2})\bigr),\quad
  c_{\varphi}=\left\{
  \begin{array}{ll}
  1&:\text{W}\\
  (\sinh\frac{-\gamma}{2})^{-1}&:\text{AW}
  \end{array}\right..
  \label{cvarphi}
\end{equation}

\medskip

First we consider a necessary condition for $X(\eta)$ giving recurrence
relations with constant coefficients.
Let us assume \eqref{XPDn} for a polynomial $X(\eta)$ of degree $L$ in $\eta$.
Applying $\hat{\mathcal{B}}_{\mathcal{D}}=\hat{\mathcal{B}}_{d_1\ldots d_M}$
\eqref{BhD_idQM} to \eqref{XPDn}, we have
\begin{align}
  &\quad\hat{\mathcal{B}}_{\mathcal{D}}
  \bigl(\check{X}(x)\check{P}_{\mathcal{D},n}(x)\bigr)
  =\sum_{k=-n}^Lr_{n,k}^{X,\mathcal{D}}\hat{\mathcal{B}}_{\mathcal{D}}
  \check{P}_{\mathcal{D},n+k}(x)
  =\sum_{k=-n}^Lr_{n,k}^{X,\mathcal{D}}
  (\mathcal{E}_{n+k}-\tilde{\mathcal{E}}_{d_M})
  \check{P}_{d_1\ldots d_{M-1},n}(\eta)\n
  &=\check{X}^{(+)}(x)\hat{\mathcal{B}}_{\mathcal{D}}
  \check{P}_{\mathcal{D},n}(x)
  -i\check{X}^{(-)}(x)\Bigl(
  \frac{ic^{\hat{\mathcal{B}}}_{\mathcal{D}}}
  {\varphi(x)\check{\Xi}_{\mathcal{D}}(x)}
  e^{\hat{\mathcal{B}}}_{\mathcal{D}}(x)
  \check{\Xi}_{d_1\ldots d_{M-1}}(x+i\tfrac{\gamma}{2})
  \check{P}_{\mathcal{D},n}(x-i\tfrac{\gamma}{2})\n
  &\qquad\qquad\qquad\qquad\qquad\qquad\qquad
  +\frac{ic^{\hat{\mathcal{B}}}_{\mathcal{D}}}
  {\varphi(x)\check{\Xi}_{\mathcal{D}}(x)}
  e^{\hat{\mathcal{B}}\,*}_{\mathcal{D}}(x)
  \check{\Xi}_{d_1\ldots d_{M-1}}(x-i\tfrac{\gamma}{2})
  \check{P}_{\mathcal{D},n}(x+i\tfrac{\gamma}{2})\Bigr)\n
  &=(\mathcal{E}_n-\tilde{\mathcal{E}}_{d_M})
  \check{X}^{(+)}(x)\check{P}_{d_1\ldots d_{M-1},n}(x)\n
  &\quad
  +\frac{c^{\hat{\mathcal{B}}}_{\mathcal{D}}}{\check{\Xi}_{\mathcal{D}}(x)}
  \frac{\check{X}^{(-)}(x)}{\varphi(x)}\Bigl(
  e^{\hat{\mathcal{B}}}_{\mathcal{D}}(x)
  \check{\Xi}_{d_1\ldots d_{M-1}}(x+i\tfrac{\gamma}{2})
  \check{P}_{\mathcal{D},n}(x-i\tfrac{\gamma}{2})\n
  &\qquad\qquad\qquad\qquad\quad
  +e^{\hat{\mathcal{B}}\,*}_{\mathcal{D}}(x)
  \check{\Xi}_{d_1\ldots d_{M-1}}(x-i\tfrac{\gamma}{2})
  \check{P}_{\mathcal{D},n}(x+i\tfrac{\gamma}{2})\Bigr),
  \label{BhDXPDn_idQM}
\end{align}
where \eqref{act_fg_idQM} and \eqref{FDPDn=} are used.
Since the expression in the first line is a polynomial in $\eta$,
the expression in the last line should be so.
By \eqref{poly} and $\eqref{eBhD}$,
$\frac{\check{X}^{(-)}(x)}{\varphi(x)}=
\frac{\check{X}(x-i\frac{\gamma}{2})-\check{X}(x+i\frac{\gamma}{2})}
{2c_{\varphi}(\eta(x-i\frac{\gamma}{2})-\eta(x+i\frac{\gamma}{2}))}$ and 
$e^{\hat{\mathcal{B}}}_{\mathcal{D}}(x)
\check{\Xi}_{d_1\ldots d_{M-1}}(x+i\tfrac{\gamma}{2})
\check{P}_{\mathcal{D},n}(x-i\tfrac{\gamma}{2})
+e^{\hat{\mathcal{B}}\,*}_{\mathcal{D}}(x)
\check{\Xi}_{d_1\ldots d_{M-1}}(x-i\tfrac{\gamma}{2})
\check{P}_{\mathcal{D},n}(x+i\tfrac{\gamma}{2})$
are polynomials in $\eta(x)$.
If the latter is not divisible by $\check{\Xi}_{\mathcal{D}}(x)$, the former
should be divisible by $\check{\Xi}_{\mathcal{D}}(x)$.

We summarize this argument as follows.
\begin{prop}
Let $X(\eta)$ be a polynomial of degree $L$ in $\eta$.
Assume \eqref{XPDn} and
\begin{equation}
  \frac{\check{X}(x-i\frac{\gamma}{2})-\check{X}(x+i\frac{\gamma}{2})}
  {\eta(x-i\frac{\gamma}{2})-\eta(x+i\frac{\gamma}{2})}
  =\check{\Xi}_{\mathcal{D}}(x)\check{Y}(x),\quad
  \text{\rm $Y(\eta)$ : a polynomial in $\eta$}.
  \label{Xcond_idQM}
\end{equation}
Then one action of $\hat{\mathcal{B}}_{\mathcal{D}}$ to the both sides of
\eqref{XPDn} keeps the polynomiality intact.
\label{propXweak_idQM}
\end{prop}
{\bf Remark}$\,$
If two polynomials in $\eta$,
$\check{\Xi}_{\mathcal{D}}(x)=\check{\Xi}_{d_1\ldots d_M}(x)$ and
$e^{\hat{\mathcal{B}}}_{\mathcal{D}}(x)
\check{\Xi}_{d_1\ldots d_{M-1}}(x+i\tfrac{\gamma}{2})
\check{P}_{\mathcal{D},n}(x-i\tfrac{\gamma}{2})
+e^{\hat{\mathcal{B}}\,*}_{\mathcal{D}}(x)
\check{\Xi}_{d_1\ldots d_{M-1}}(x-i\tfrac{\gamma}{2})
\check{P}_{\mathcal{D},n}(x+i\tfrac{\gamma}{2})$,
have no common roots for some $n$,
the polynomial of degree $L$ in $\eta$, $X(\eta)$, satisfying \eqref{XPDn}
should satisfy \eqref{Xcond_idQM} for some polynomial $Y(\eta)$.

By taking the constant term of $X(\eta)$ as $X(0)=0$, the condition for the
candidate of $X(\eta)$ \eqref{Xcond_idQM} gives
\begin{equation}
  X(\eta)=I[\Xi_{\mathcal{D}}Y](\eta),\quad
  \text{deg}\,X(\eta)=L=\ell_{\mathcal{D}}+\text{deg}\,Y(\eta)+1.
  \label{X=int_idQM}
\end{equation}
The minimal degree candidate of $X(\eta)$, which corresponds to $Y(\eta)=1$, is
\begin{equation}
  X_{\text{min}}(\eta)=I[\Xi_{\mathcal{D}}](\eta),\quad
  \text{deg}\,X_{\text{min}}(\eta)=\ell_{\mathcal{D}}+1.
  \label{Xmin_idQM}
\end{equation}

Based on these properties we present our another main result.
Like as \S\,\ref{sec:RRoQM},
we conjecture that this candidate $X(\eta)$ \eqref{X=int_idQM} actually gives
recurrence relations with constant coefficients.
\begin{conj}
For any polynomial $Y(\eta)$, we take $X(\eta)$ as \eqref{X=int_idQM}.
Then the multi-indexed Wilson and Askey-Wilson polynomials
$P_{\mathcal{D},n}(\eta)$ satisfy $1+2L$ term recurrence relations with
constant coefficients \eqref{XP}.
\label{conj_idQM}
\end{conj}
{\bf Remark 1}$\,$
If two polynomials in $\eta$,
$\check{\Xi}_{\mathcal{D}}(x)=\check{\Xi}_{d_1\ldots d_M}(x)$ and
$e^{\hat{\mathcal{B}}}_{\mathcal{D}}(x)
\check{\Xi}_{d_1\ldots d_{M-1}}(x+i\tfrac{\gamma}{2})
\check{P}_{\mathcal{D},n}(x-i\tfrac{\gamma}{2})
+e^{\hat{\mathcal{B}}\,*}_{\mathcal{D}}(x)
\check{\Xi}_{d_1\ldots d_{M-1}}(x-i\tfrac{\gamma}{2})
\check{P}_{\mathcal{D},n}(x+i\tfrac{\gamma}{2})$,
have no common roots for some $n$, this conjecture exhausts all
possible $X(\eta)$ giving recurrence relations with constant coefficients, and
the minimal degree choice $X(\eta)=X_{\text{min}}(\eta)$ \eqref{Xmin_idQM} gives
$3+2\ell_{\mathcal{D}}$ term recurrence relations.\\
{\bf Remark 2}$\,$
See Remark 2 and 3 below Conjecture\,\ref{conj_oQM}.\\
{\bf Remark 3}$\,$
By \eqref{poly}, for any polynomial $p(\eta)$,
$\check{Y}(x)
=\frac{\check{\Xi}_{\mathcal{D}}(x-i\gamma)\check{p}(x-i\frac{\gamma}{2})
-\check{\Xi}_{\mathcal{D}}(x+i\gamma)\check{p}(x+i\frac{\gamma}{2})}
{\eta(x-i\frac{\gamma}{2})-\eta(x+i\frac{\gamma}{2})}$ is also a polynomial
in $\eta$.
This $Y(\eta)$ gives
$\check{X}(x)
=\check{\Xi}_{\mathcal{D}}(x-i\tfrac{\gamma}{2})
\check{\Xi}_{\mathcal{D}}(x+i\tfrac{\gamma}{2})\check{p}(x)$, which corresponds
to Remark 4 below Conjecture\,\ref{conj_oQM}.\\
{\bf Remark 4}$\,$
See Remark 5 below Conjecture\,\ref{conj_oQM}.\\
{\bf Remark 5}$\,$
For $M=1$ we can prove this conjecture.
Since we have Proposition\,\ref{rrXPDn}, it is sufficient to show that
$\hat{\mathcal{B}}_{d_1}\hat{\mathcal{B}}_{d_1d_2}\cdots
\hat{\mathcal{B}}_{d_1\ldots d_M}(XP_{\mathcal{D},n})$
is a polynomial in $\eta$.\\
\underline{$M=1$}
{}From \eqref{BhDXPDn_idQM} we have
\begin{align*}
  &\quad\hat{\mathcal{B}}_{d_1}\bigl(\check{X}(x)\check{P}_{d_1,n}(x)\bigr)\n
  &=(\mathcal{E}_n-\tilde{\mathcal{E}}_{d_1})\check{X}^{(+)}(x)\check{P}_n(x)
  +\frac{c^{\hat{\mathcal{B}}}_{d_1}}{2c_{\varphi}}\check{Y}(x)
  \bigl(e^{\hat{\mathcal{B}}}_{d_1}(x)\check{P}_{d_1,n}(x-i\tfrac{\gamma}{2})
  +e^{\hat{\mathcal{B}}\,*}_{d_1}(x)
  \check{P}_{d_1,n}(x+i\tfrac{\gamma}{2})\bigr).
\end{align*}
This is a polynomial in $\eta$. Thus $M=1$ case is proved.
\hfill\fbox{}

\section{Summary and Comments}
\label{sec:summary}

In addition to $3+2M$ term recurrence relations with variable dependent
coefficients presented in \cite{rrmiop}, we have presented (conjectures of)
the recurrence relations with constant coefficients for the multi-indexed
orthogonal polynomials of Laguerre, Jacobi, Wilson and Askey-Wilson types,
Conjecture\,\ref{conj_oQM} and Conjecture\,\ref{conj_idQM}.
Since $Y(\eta)$ is arbitrary, we obtain infinitely many recurrence relations.
However not all of them are independent.
The most important one is the minimal degree one $X_{\text{min}}(\eta)$
\eqref{Xmin} or \eqref{Xmin_idQM}, which gives $3+2\ell_{\mathcal{D}}$
term recurrence relations. Here $\ell_{\mathcal{D}}$ ($\geq M$) is the degree
of the lowest member polynomial $P_{\mathcal{D},0}(\eta)$.
For this case, the coefficients
$r_{n,k}^{X_{\text{min}},\mathcal{D}}$ may have nice forms.
Both derivations given in \cite{rrmiop} and present paper are based on
multi-step Darboux transformations. Although we have discussed Laguerre,
Jacobi, Wilson and Askey-Wilson cases, the method is applicable to
the multi-indexed ($q$-)Racah polynomials (which correspond to the case (1)
$\mathcal{I}=\{0,1,\ldots,\ell-1\}$) and various case (2) polynomials.
The results in \cite{stz} and \cite{mt} (type $\I$ and $\II$) correspond to
special cases of our results.
In \cite{mt}, type $\III$ case is also studied, which corresponds to case (2).
In \cite{duran}, exceptional Charlier, Meixner, Hermite and Laguerre
polynomials are studied and some recurrence relations, which have minimal
order (minimalness is stated as a conjecture), are proved by a different
method from our paper.
The exceptional Laguerre polynomials in \cite{duran} are labeled by
the two sets, $F_1=\{f_1,\ldots,f_{k_1}\}$ (labels of eigenstates)
and $F_2=\{f'_1,\ldots,f'_{k_2}\}$ (labels of type $\I$ virtual states).
For $F_1=\emptyset$, they correspond to special cases of our multi-indexed
Laguerre polynomials with $\mathcal{D}=\{d_1^{\I},\ldots,d_M^{\I}\}=F_2$
(no type $\II$), for which existence of (minimal degree) recurrence relations
are proved \cite{duran}.
For $F_1\neq\emptyset$, they correspond to case (2). 
For general case (2), the referee suggests that the minimal degree is equal
to the number of missing degrees.
It is an interesting problem to show this suggestion.
We hope that Conjecture\,\ref{conj_oQM} and \ref{conj_idQM} will be
proved in the near future.

Multi-indexed orthogonal polynomials of Laguerre, Jacobi, Wilson and
Askey-Wilson types are labeled by an index set $\mathcal{D}$ but different
index sets may give the same multi-indexed orthogonal polynomials,
$P_{\mathcal{D},n}(\eta;\bm{\lambda})
\propto P_{\mathcal{D}',n}(\eta;\bm{\lambda}')$ \cite{equiv_miop}.
For example, $\mathcal{D}_1=\{1^{\II},3^{\II},4^{\II},5^{\II},8^{\II}\}$
with $\bm{\lambda}$, $\mathcal{D}_2=\{1^{\I},2^{\I},6^{\I},8^{\I}\}$
with $\bm{\lambda}-9\tilde{\bm{\delta}}_{\I}$ and
$\mathcal{D}_3=\{3^{\I},5^{\I},2^{\II}\}$ with
$\bm{\lambda}-6\tilde{\bm{\delta}}_{\I}$ give the same multi-indexed
orthogonal polynomials,
$P_{\mathcal{D}_1,n}(\eta;\bm{\lambda})\propto
P_{\mathcal{D}_2,n}(\eta;\bm{\lambda}-9\tilde{\bm{\delta}}_{\I})\propto
P_{\mathcal{D}_3,n}(\eta;\bm{\lambda}-6\tilde{\bm{\delta}}_{\I})$.
The $3+2M$ term recurrence relations given in \cite{rrmiop} states that
these polynomials $P_{\mathcal{D}_1,n}(\eta;\bm{\lambda})$,
$P_{\mathcal{D}_2,n}(\eta;\bm{\lambda}-9\tilde{\bm{\delta}}_{\I})$ and
$P_{\mathcal{D}_3,n}(\eta;\bm{\lambda}-6\tilde{\bm{\delta}}_{\I})$ satisfy
13, 11 and 9 term recurrence relations with variable dependent coefficients,
respectively. But the above equivalence implies that all of them satisfy
9 term recurrence relations with variable dependent coefficients.
On the other hand, the degrees of lowest members $P_{\mathcal{D},0}(\eta)$
are $\ell_{\mathcal{D}_1}=\ell_{\mathcal{D}_2}=\ell_{\mathcal{D}_3}=11$, and
for each case the minimal degree polynomial $X_{\text{min}}(\eta)$ gives 25
term recurrence relations with constant coefficients.
The above equivalence, which gives 
$\Xi_{\mathcal{D}_1}(\eta;\bm{\lambda})\propto
\Xi_{\mathcal{D}_2}(\eta;\bm{\lambda}-9\tilde{\bm{\delta}}_{\I})\propto
\Xi_{\mathcal{D}_3}(\eta;\bm{\lambda}-6\tilde{\bm{\delta}}_{\I})$,
implies that these three 25 term recurrence relations are essentially same.

The $3+2M$ term recurrence relations with variable dependent coefficients
can be used to calculate the multi-indexed orthogonal polynomials effectively
and it needs $M+1$ initial data $P_{\mathcal{D},n}(\eta)$ ($n=0,1,\dots,M$)
\cite{rrmiop}.
The simplest recurrence relations with constant coefficients corresponding
to $X_{\text{min}}(\eta)$ has $3+2\ell_{\mathcal{D}}$ terms and it needs
$\ell_{\mathcal{D}}+1$ initial data.
The difference of $M$ and $\ell_{\mathcal{D}}$,
$\ell_{\mathcal{D}}-M=\sum_{j=1}^M(d_j-1)+2M_{\I}M_{\II}$, becomes large for
large $d_j$.
In order to calculate the multi-indexed orthogonal polynomials by using
recurrence relations, the $3+2M$ term recurrence relations with variable
dependent coefficients are useful.
On the other hand, in order to study bispectral properties etc., recurrence
relations with constant coefficients are needed.

We hope that the recurrence relations with constants coefficients obtained
in this paper will be used as a starting point for theoretical developments
of various problems involving bispectrality, generalizations of the Jacobi
matrix, spectral theory, etc.

\section*{Acknowledgments}

I thank the organizers and participants of the workshop ``Exceptional
orthogonal polynomials and exact solutions in Mathematical Physics''
at Segovia (Spain), 7-12 September 2014.
This workshop was exciting and I learned importance of the recurrence
relations with constant coefficients.
I thank R.\,Sasaki for discussion and reading of the manuscript,
and S.\,Tsujimoto and H.\,Miki for discussion.
I am supported in part by Grant-in-Aid for Scientific Research
from the Ministry of Education, Culture, Sports, Science and Technology
(MEXT), No.25400395.

\bigskip
\appendix
\section{Some Formulas}
\label{sec:A}

The notation and fundamental formulas of the multi-indexed orthogonal
polynomials of Laguerre, Jacobi, Wilson and Askey-Wilson types are found in
\cite{rrmiop}. See also \cite{os27}.
In this appendix we present other basic formulas.
See footnote in \S\,\ref{sec:method}.
We write parameter ($\bm{\lambda}$) dependence explicitly.

\subsection{Multi-indexed Laguerre and Jacobi polynomials}
\label{sec:A_LJ}

Explicit forms of the operators \eqref{FhD=} are
\begin{align}
  &\hat{\mathcal{F}}_{d_1\ldots d_s}(\bm{\lambda})
  =\cF^{-1}\frac{\Xi_{d_1\ldots d_s}(\eta;\bm{\lambda})}
  {\Xi_{d_1\ldots d_{s-1}}(\eta;\bm{\lambda})}
  \biggl(\cF e^{\hat{\mathcal{F}}}_{d_1\ldots d_s}(\eta)
  \Bigl(\frac{d}{d\eta}
  -\frac{\partial_{\eta}\Xi_{d_1\ldots d_s}(\eta;\bm{\lambda})}
  {\Xi_{d_1\ldots d_s}(\eta;\bm{\lambda})}\Bigr)
  +\tilde{e}^{\hat{\mathcal{F}}}_{d_1\ldots d_s}(\bm{\lambda})\biggr),
  \label{FhD}\\
  &\hat{\mathcal{B}}_{d_1\ldots d_s}(\bm{\lambda})
  =\cF\frac{\Xi_{d_1\ldots d_{s-1}}(\eta;\bm{\lambda})}
  {\Xi_{d_1\ldots d_s}(\eta;\bm{\lambda})}
  \biggl(\cF e^{\hat{\mathcal{B}}}_{d_1\ldots d_s}(\eta)
  \Bigl(-\frac{d}{d\eta}
  +\frac{\partial_{\eta}\Xi_{d_1\ldots d_{s-1}}(\eta;\bm{\lambda})}
  {\Xi_{d_1\ldots d_{s-1}}(\eta;\bm{\lambda})}\Bigr)
  +\tilde{e}^{\hat{\mathcal{B}}}_{d_1\ldots d_s}(\bm{\lambda})\biggr),
  \label{BhD}
\end{align}
where $e^{\hat{\mathcal{F}}}_{d_1\ldots d_s}(\eta)$,
$e^{\hat{\mathcal{B}}}_{d_1\ldots d_s}(\eta)$,
$\tilde{e}^{\hat{\mathcal{F}}}_{d_1\ldots d_s}(\bm{\lambda})$ and
$\tilde{e}^{\hat{\mathcal{B}}}_{d_1\ldots d_s}(\bm{\lambda})$ are given by
\begin{align}
  \text{L}:\quad
  &e^{\hat{\mathcal{F}}}_{d_1\ldots d_s}(\eta)
  =\left\{\begin{array}{ll}
  1&:d_s=d_s^{\I}\\[2pt]
  \eta&:d_s=d_s^{\II}
  \end{array}\right.,\quad
  e^{\hat{\mathcal{B}}}_{d_1\ldots d_s}(\eta)
  =\left\{\begin{array}{ll}
  \eta&:d_s=d_s^{\I}\\[2pt]
  1&:d_s=d_s^{\II}
  \end{array}\right.,\n
  &\tilde{e}^{\hat{\mathcal{F}}}_{d_1\ldots d_s}(\bm{\lambda})
  =\left\{\begin{array}{ll}
  -2&:d_s=d_s^{\I}\\[2pt]
  2(g+s_{\I}-s_{\II})+1&:d_s=d_s^{\II}
  \end{array}\right.,\n
  &\tilde{e}^{\hat{\mathcal{B}}}_{d_1\ldots d_s}(\bm{\lambda})
  =\left\{\begin{array}{ll}
  -2(g+s_{\I}-s_{\II})+1&:d_s=d_s^{\I}\\[2pt]
  2&:d_s=d_s^{\II}
  \end{array}\right.,\\
  \text{J}:\quad
  &e^{\hat{\mathcal{F}}}_{d_1\ldots d_s}(\eta)
  =\left\{\begin{array}{ll}
  \frac{1+\eta}{2}&:d_s=d_s^{\I}\\[2pt]
  \frac{1-\eta}{2}&:d_s=d_s^{\II}
  \end{array}\right.,\quad
  e^{\hat{\mathcal{B}}}_{d_1\ldots d_s}(\eta)
  =\left\{\begin{array}{ll}
  \frac{1-\eta}{2}&:d_s=d_s^{\I}\\[2pt]
  \frac{1+\eta}{2}&:d_s=d_s^{\II}
  \end{array}\right.,\n
  &\tilde{e}^{\hat{\mathcal{F}}}_{d_1\ldots d_s}(\bm{\lambda})
  =\left\{\begin{array}{ll}
  -2(h+s_{\II}-s_{\I})-1&:d_s=d_s^{\I}\\[2pt]
  2(g+s_{\I}-s_{\II})+1&:d_s=d_s^{\II}
  \end{array}\right.,\n
  &\tilde{e}^{\hat{\mathcal{B}}}_{d_1\ldots d_s}(\bm{\lambda})
  =\left\{\begin{array}{ll}
  -2(g+s_{\I}-s_{\II})+1&:d_s=d_s^{\I}\\[2pt]
  2(h+s_{\II}-s_{\I})-1&:d_s=d_s^{\II}
  \end{array}\right..
\end{align}
By \eqref{FhD}--\eqref{BhD}, eqs.\,\eqref{FDPDn=} are
\begin{align}
  &\quad P_{d_1\ldots d_s,n}(\eta;\bm{\lambda})\n
  &=\frac{1}{\Xi_{d_1\dots d_{s-1}}(\eta;\bm{\lambda})}\Bigl(
  e^{\hat{\mathcal{F}}}_{d_1\ldots d_s}(\eta)\bigl(
  \Xi_{d_1\ldots d_s}(\eta;\bm{\lambda})
  \partial_{\eta}P_{d_1\ldots d_{s-1},n}(\eta;\bm{\lambda})
  -\partial_{\eta}\Xi_{d_1\ldots d_s}(\eta;\bm{\lambda})
  P_{d_1\ldots d_{s-1},n}(\eta;\bm{\lambda})\bigr)\n
  &\qquad\qquad\qquad\qquad
  +\cF^{-1}\tilde{e}^{\hat{\mathcal{F}}}_{d_1\ldots d_s}(\bm{\lambda})
  \Xi_{d_1\ldots d_s}(\eta;\bm{\lambda})
  P_{d_1\ldots d_{s-1},n}(\eta;\bm{\lambda})\Bigr),
  \label{PDnF}\\
  &\quad\big(\mathcal{E}_n(\bm{\lambda})
  -\tilde{\mathcal{E}}_{d_s}(\bm{\lambda})\bigr)
  P_{d_1\ldots d_{s-1},n}(\eta;\bm{\lambda})\n
  &=\frac{\cF^2}{\Xi_{d_1\dots d_s}(\eta;\bm{\lambda})}
  \Bigl(e^{\hat{\mathcal{B}}}_{d_1\ldots d_s}(\eta)\bigl(
  -\Xi_{d_1\ldots d_{s-1}}(\eta;\bm{\lambda})
  \partial_{\eta}P_{d_1\ldots d_s,n}(\eta;\bm{\lambda})
  +\partial_{\eta}\Xi_{d_1\ldots d_{s-1}}(\eta;\bm{\lambda})
  P_{d_1\ldots d_s,n}(\eta;\bm{\lambda})\bigr)\n
  &\qquad\qquad\qquad\ \quad
  +\cF^{-1}\tilde{e}^{\hat{\mathcal{B}}}_{d_1\ldots d_s}(\bm{\lambda})
  \Xi_{d_1\ldots d_{s-1}}(\eta;\bm{\lambda})
  P_{d_1\ldots d_s,n}(\eta;\bm{\lambda})\Bigr).
  \label{PDnB}
\end{align}

\subsection{Multi-indexed Wilson and Askey-Wilson polynomials}
\label{sec:A_WAW}

We restrict the parameters: $\{a_1^*,a_2^*\}=\{a_1,a_2\}$ (as a set) and
$\{a_3^*,a_4^*\}=\{a_3,a_4\}$ (as a set).

Explicit form of the potential function
$\hat{V}_{d_1\ldots d_s}(x;\bm{\lambda})$ is
\begin{align}
  \hat{V}_{d_1\ldots d_s}(x;\bm{\lambda})
  &=\frac{\check{\Xi}_{d_1\ldots d_{s-1}}(x+i\frac{\gamma}{2};\bm{\lambda})}
  {\check{\Xi}_{d_1\ldots d_{s-1}}(x-i\frac{\gamma}{2};\bm{\lambda})}
  \frac{\check{\Xi}_{d_1\ldots d_s}(x-i\gamma;\bm{\lambda})}
  {\check{\Xi}_{d_1\ldots d_s}(x;\bm{\lambda})}\n
  &\quad\times\left\{
  \begin{array}{ll}
  \kappa^{s_{\I}-s_{\II}-1}\alpha^{\I}(\bm{\lambda})
  V\bigl(x;\mathfrak{t}^{\I}(\bm{\lambda}^{[s_{\I}-1,s_{\II}]})\bigr)
  &:d_s=d_s^{\I}\\[2pt]
  \kappa^{s_{\II}-s_{\I}-1}\alpha^{\II}(\bm{\lambda})
  V\bigl(x;\mathfrak{t}^{\II}(\bm{\lambda}^{[s_{\I},s_{\II}-1]})\bigr)
  &:d_s=d_s^{\II}
  \end{array}\right..
\end{align}
Explicit forms of the operators \eqref{FhD=} are
\begin{align}
  &\hat{\mathcal{F}}_{d_1\ldots d_s}(\bm{\lambda})
  =\frac{ic^{\hat{\mathcal{F}}}_{d_1\ldots d_s}(\bm{\lambda})}
  {\varphi(x)\check{\Xi}_{d_1\ldots d_{s-1}}(x;\bm{\lambda})}\Bigl(
  e^{\hat{\mathcal{F}}}_{d_1\ldots d_s}(x;\bm{\lambda})
  \check{\Xi}_{d_1\ldots d_s}(x+i\tfrac{\gamma}{2};\bm{\lambda})
  e^{\frac{\gamma}{2}p}\n
  &\qquad\qquad\qquad\qquad\qquad\qquad\ \quad
  -e^{\hat{\mathcal{F}}\,*}_{d_1\ldots d_s}(x;\bm{\lambda})
  \check{\Xi}_{d_1\ldots d_s}(x-i\tfrac{\gamma}{2};\bm{\lambda})
  e^{-\frac{\gamma}{2}p}\Bigr),
  \label{FhD_idQM}\\
  &\hat{\mathcal{B}}_{d_1\ldots d_s}(\bm{\lambda})
  =\frac{ic^{\hat{\mathcal{B}}}_{d_1\ldots d_s}(\bm{\lambda})}
  {\varphi(x)\check{\Xi}_{d_1\ldots d_s}(x;\bm{\lambda})}\Bigl(
  e^{\hat{\mathcal{B}}}_{d_1\ldots d_s}(x;\bm{\lambda})
  \check{\Xi}_{d_1\ldots d_{s-1}}(x+i\tfrac{\gamma}{2};\bm{\lambda})
  e^{\frac{\gamma}{2}p}\n
  &\qquad\qquad\qquad\qquad\qquad\ \qquad
  -e^{\hat{\mathcal{B}}\,*}_{d_1\ldots d_s}(x;\bm{\lambda})
  \check{\Xi}_{d_1\ldots d_{s-1}}(x-i\tfrac{\gamma}{2};\bm{\lambda})
  e^{-\frac{\gamma}{2}p}\Bigr),
  \label{BhD_idQM}
\end{align}
where $c^{\hat{\mathcal{F}}}_{d_1\ldots d_s}(\bm{\lambda})$,
$c^{\hat{\mathcal{B}}}_{d_1\ldots d_s}(\bm{\lambda})$,
$e^{\hat{\mathcal{F}}}_{d_1\ldots d_s}(x;\bm{\lambda})$ and
$e^{\hat{\mathcal{B}}}_{d_1\ldots d_s}(x;\bm{\lambda})$ are given by
\begin{align}
  &c^{\hat{\mathcal{F}}}_{d_1\ldots d_s}(\bm{\lambda})^{-1}
  =c^{\hat{\mathcal{B}}}_{d_1\ldots d_s}(\bm{\lambda})
  =\left\{\begin{array}{ll}
  \kappa^{\frac{s-1}{2}+s_{\II}}\alpha^{\I}(\bm{\lambda})^{\frac12}
  &:d_s=d_s^{\I}\\
  \kappa^{\frac{s-1}{2}+s_{\I}}\alpha^{\II}(\bm{\lambda})^{\frac12}
  &:d_s=d_s^{\II}
  \end{array}\right.,
  \label{cBhD}\\
  &e^{\hat{\mathcal{F}}}_{d_1\ldots d_s}(x;\bm{\lambda})
  =\left\{\begin{array}{ll}
  v_1(x;\bm{\lambda}^{[s_{\I},s_{\II}]})&:d_s=d_s^{\I}\\
  v_2(x;\bm{\lambda}^{[s_{\I},s_{\II}]})&:d_s=d_s^{\II}
  \end{array}\right.,
  \label{eFhD}\\
  &e^{\hat{\mathcal{B}}}_{d_1\ldots d_s}(x;\bm{\lambda})
  =\left\{\begin{array}{ll}
  v_2(x;\bm{\lambda}^{[s_{\I}-1,s_{\II}]})&:d_s=d_s^{\I}\\
  v_1(x;\bm{\lambda}^{[s_{\I},s_{\II}-1]})&:d_s=d_s^{\II}
  \end{array}\right..
  \label{eBhD}
\end{align}
Here $v_1(x;\bm{\lambda})$ and $v_2(x;\bm{\lambda})$ are given in \cite{os27}:
\begin{equation}
  v_1(x;\bm{\lambda})=\left\{
  \begin{array}{ll}
  \!\!\prod_{j=1}^2(a_j+ix)&\!\!:\text{W}\\[2pt]
  \!\!e^{-ix}\prod_{j=1}^2(1-a_je^{ix})&\!\!:\text{AW}
  \end{array}\right.\!\!\!,
  \ \ v_2(x;\bm{\lambda})=\left\{
  \begin{array}{ll}
  \!\!\prod_{j=3}^4(a_j+ix)&\!\!:\text{W}\\[2pt]
  \!\!e^{-ix}\prod_{j=3}^4(1-a_je^{ix})&\!\!:\text{AW}
  \end{array}\right.\!\!\!.\!\!\!
  \label{v1v2}
\end{equation}
Note that $v_1^*(x;\bm{\lambda})=v_1(-x;\bm{\lambda})$ and
$v_2^*(x;\bm{\lambda})=v_2(-x;\bm{\lambda})$.
By \eqref{FhD_idQM}--\eqref{BhD_idQM}, eqs.\,\eqref{FDPDn=} are
\begin{align}
  &\check{P}_{d_1\dots d_s,n}(x;\bm{\lambda})\n
  &\quad=\frac{ic^{\hat{\mathcal{F}}}_{d_1\ldots d_s}(\bm{\lambda})}
  {\varphi(x)\check{\Xi}_{d_1\ldots d_{s-1}}(x;\bm{\lambda})}\Bigl(
  e^{\hat{\mathcal{F}}}_{d_1\ldots d_s}(x;\bm{\lambda})
  \check{\Xi}_{d_1\ldots d_s}(x+i\tfrac{\gamma}{2};\bm{\lambda})
  P_{d_1\ldots d_{s-1},n}(x-i\tfrac{\gamma}{2};\bm{\lambda})\n
  &\qquad\qquad\qquad\qquad\qquad
  -e^{\hat{\mathcal{F}}\,*}_{d_1\ldots d_s}(x;\bm{\lambda})
  \check{\Xi}_{d_1\ldots d_s}(x-i\tfrac{\gamma}{2};\bm{\lambda})
  P_{d_1\ldots d_{s-1},n}(x+i\tfrac{\gamma}{2};\bm{\lambda})\Bigr),
  \label{PDnF_idQM}\\
  &\bigl(\mathcal{E}_n(\bm{\lambda})
  -\tilde{\mathcal{E}}_{d_s}(\bm{\lambda})\bigr)
  P_{d_1\ldots d_{s-1},n}(x;\bm{\lambda})\n
  &\quad=\frac{ic^{\hat{\mathcal{B}}}_{d_1\ldots d_s}(\bm{\lambda})}
  {\varphi(x)\check{\Xi}_{d_1\ldots d_s}(x;\bm{\lambda})}\Bigl(
  e^{\hat{\mathcal{B}}}_{d_1\ldots d_s}(x;\bm{\lambda})
  \check{\Xi}_{d_1\ldots d_{s-1}}(x+i\tfrac{\gamma}{2};\bm{\lambda})
  P_{d_1\ldots d_s,n}(x-i\tfrac{\gamma}{2};\bm{\lambda})\n
  &\qquad\qquad\qquad\qquad\,\quad
  -e^{\hat{\mathcal{B}}\,*}_{d_1\ldots d_s}(x;\bm{\lambda})
  \check{\Xi}_{d_1\ldots d_{s-1}}(x-i\tfrac{\gamma}{2};\bm{\lambda})
  P_{d_1\ldots d_s,n}(x+i\tfrac{\gamma}{2};\bm{\lambda})\Bigr).
  \label{PDnB_idQM}
\end{align}

\section{Some Examples}
\label{sec:B}

For illustration, we present some examples of the coefficients
$r_{n,k}^{X,\mathcal{D}}$ of the recurrence relations \eqref{XP} for
$X(\eta)=X_{\text{min}}(\eta)$ and small $d_j$.

\subsection{Multi-indexed Laguerre polynomials}
\label{sec:B_L}

\noindent
\underline{Ex.1} $\mathcal{D}=\{1^{\I}\}$ : 5-term recurrence relations
\begin{align}
  X(\eta)&=X_{\text{min}}(\eta)=\tfrac12\eta(\eta+2g+1),\n
  r_{n,2}^{X,\mathcal{D}}&=\tfrac12(n+1)(n+2),\quad
  r_{n,-2}^{X,\mathcal{D}}=\tfrac18(2g+2n-3)(2g+2n+3),\n
  r_{n,1}^{X,\mathcal{D}}&=-(n+1)(2g+2n+3),\quad
  r_{n,-1}^{X,\mathcal{D}}=-\tfrac12(2g+2n-1)(2g+2n+3),
  \label{Ex1_L}\\
  r_{n,0}^{X,\mathcal{D}}
  &=\tfrac18\bigl(24n^2+4(10g+11)n+(2g+1)(6g+13)\bigr),
  \nonumber
\end{align}
\underline{Ex.2} $\mathcal{D}=\{1^{\I},2^{\I}\}$ : 7-term recurrence relations
\begin{align}
  X(\eta)&=X_{\text{min}}(\eta)=
  \tfrac{1}{24}\eta\bigl(4\eta^2+6(2g+1)\eta+3(2g+1)(2g+3)\bigr),\n
  r_{n,3}^{X,\mathcal{D}}&=-\tfrac16(n+1)_3\,,\quad
  r_{n,-3}^{X,\mathcal{D}}=-\tfrac{1}{12}(2g+2n-5)(g+n+\tfrac32)_2\,,\n
  r_{n,2}^{X,\mathcal{D}}&=\tfrac12(n+1)_2(2g+2n+5),\quad
  r_{n,-2}^{X,\mathcal{D}}=\tfrac12(2g+2n-3)(g+n+\tfrac32)_2\,,\n
  r_{n,1}^{X,\mathcal{D}}&=-\tfrac14(n+1)(2g+2n+3)(4g+5n+12),
  \label{Ex2_L}\\
  r_{n,-1}^{X,\mathcal{D}}&=-\tfrac18(2g+2n-1)(2g+2n+5)(4g+5n+7),\n
  r_{n,0}^{X,\mathcal{D}}&=\tfrac{1}{48}\bigl(
  160n^3+96(4g+7)n^2+8(36g^2+132g+97)n+(2g+1)(2g+5)(14g+45)\bigr),
  \nonumber
\end{align}
\underline{Ex.3} $\mathcal{D}=\{1^{\I},1^{\II}\}$ : 9-term recurrence relations
\begin{align}
  X(\eta)&=X_{\text{min}}(\eta)=
  \tfrac18\eta\bigl(2\eta^3+4(2g-1)\eta^2+3(2g-3)(2g+1)\eta
  +(2g-3)(2g-1)(2g+1)\bigr),\n
  r_{n,4}^{X,\mathcal{D}}&=\frac{(n+1)_4(2g+2n-3)}{4(2g+2n+5)},\quad
  r_{n,-4}^{X,\mathcal{D}}=\tfrac{1}{16}(2g+2n-7)(g+n-\tfrac32)_2(2g+2n+3),\n
  r_{n,3}^{X,\mathcal{D}}&=-(n+1)_3(2g+2n-3),\quad
  r_{n,-3}^{X,\mathcal{D}}=-(g+n-\tfrac52)_3(2g+2n+3),\n
  r_{n,2}^{X,\mathcal{D}}&=\frac{(n+1)_2(2g+2n-3)}{4(2g+2n+1)}
  \bigl(28n^2+2(26g+29)n+3(2g+1)(4g+7)\bigr),\n
  r_{n,-2}^{X,\mathcal{D}}&=\tfrac{1}{16}(2g+2n-3)(2g+2n+3)
  \bigl(28n^2+2(26g-27)n+24g^2-50g+17\bigr),\\
  r_{n,1}^{X,\mathcal{D}}&=-\tfrac12(n+1)(2g+2n-3)(2g+2n+3)(4g+7n+5),\n
  r_{n,-1}^{X,\mathcal{D}}&=-(g+n-\tfrac32)_2(2g+2n+3)(4g+7n-2),\n
  r_{n,0}^{X,\mathcal{D}}&=\tfrac{1}{64}\bigl(
  1120n^4+160(22g+3)n^3+8(492g^2+168g-299)n^2\n
  &\qquad
  +8(224g^3+156g^2-328g-135)n+(2g-3)(2g+1)(6g+5)(10g+19)\bigr).
  \nonumber
\end{align}
Note that we have equivalences \cite{equiv_miop},
\begin{align}
  P_{\{1^{\I},2^{\I}\},n}(\eta;g)
  &=\frac{1}{g+n+\frac12}
  P_{\{2^{\II}\},n}(\eta;g+3),\\
  P_{\{1^{\I},1^{\II}\},n}(\eta;g)
  &=-3(g+n-\tfrac32)
  P_{\{1^{\I},3^{\I}\},n}(\eta;g-2).
\end{align}
Recurrence relations for $P_{\{1^{\I}\},n}(\eta;g)$ were given in
\cite{mt,duran} and those for $P_{\{2^{\II}\},n}(\eta;g)$ were given in
\cite{mt}.

\subsection{Multi-indexed Jacobi polynomials}
\label{sec:B_J}

We set $a=g+h$ and $b=g-h$.\\
\underline{Ex.1} $\mathcal{D}=\{1^{\I}\}$ : 5-term recurrence relations
\begin{align}
  X(\eta)&=X_{\text{min}}(\eta)=
  \tfrac14\eta\bigl((b+2)\eta+2(a-1)\bigr),\n
  r_{n,2}^{X,\mathcal{D}}&=
  \frac{(n+1)_2(b+2)(a+n)_2(2h+2n-3)}{(a+2n)_4(2h+2n+1)},\n
  r_{n,-2}^{X,\mathcal{D}}&=
  \frac{(b+2)(2g+2n-3)(2g+2n+3)(h+n-\tfrac32)_2}{4(a+2n-3)_4},\n
  r_{n,1}^{X,\mathcal{D}}&=
  \frac{(n+1)(a-1)(a+n)(2g+2n+3)(2h+2n-3)}{(a+2n-1)_3(a+2n+3)},
  \label{Ex1_J}\\
  r_{n,-1}^{X,\mathcal{D}}&=
  \frac{(a-1)(2g+2n-1)(2g+2n+3)(h+n-\tfrac32)_2}{(a+2n-3)(a+2n-1)_3},\n
  r_{n,0}^{X,\mathcal{D}}&=
  \frac{b+2}{4(a+2n-2)_2(a+2n+1)_2}\Bigl(
  -b(b+4)\bigl(2n(a+n)-(a-2)(a-1)\bigr)\n
  &\qquad\qquad
  +(a+2n-1)(a+2n+1)\bigl(2n(a+n)-(a-2)(2a-1)\bigr)\Bigr),
  \nonumber
\end{align}
\underline{Ex.2} $\mathcal{D}=\{1^{\I},2^{\I}\}$ : 7-term recurrence relations
\begin{align}
  X(\eta)&=X_{\text{min}}(\eta)=
  \tfrac{1}{48}(b+4)\eta\bigl((b+2)(b+3)\eta^2+3(b+3)(a-1)\eta
  +3(a^2-2a+b+3)\bigr),\n
  r_{n,3}^{X,\mathcal{D}}&=
  \frac{(n+1)_3(b+2)_3(a+n)_3(h+n-\tfrac52)_2}{6(a+2n)_6(h+n+\tfrac12)_2},\n
  r_{n,-3}^{X,\mathcal{D}}&=
  \frac{(b+2)_3(2g+2n-5)(g+n+\tfrac32)_2(h+n-\tfrac52)_3}{12(a+2n-5)_6},\n
  r_{n,2}^{X,\mathcal{D}}&=
  \frac{(n+1)_2(b+3)_2(a-1)(a+n)_2(2g+2n+5)(h+n-\tfrac52)_2}
  {(a+2n-1)_5(a+2n+5)(2h+2n+1)},\n
  r_{n,-2}^{X,\mathcal{D}}&=
  \frac{(b+3)_2(a-1)(2g+2n-3)(g+n+\tfrac32)_2(h+n-\tfrac52)_3}
  {2(a+2n-5)(a+2n-3)_5},\n
  r_{n,1}^{X,\mathcal{D}}&=
  \frac{(b+4)(a+n)(2g+2n+3)(2h+2n-5)}{8(a+2n-2)_4(a+2n+3)_2}\n
  &\quad\times\Bigl(
   b(b+9)(n+1)\bigl(n(n+a+1)-(a-2)_2\bigr)\n
  &\qquad
   +(n+1)\bigl(2(9-4a+2a^2)n(n+a+1)+(a-3)_3(a+6)\bigr)\Bigr),
  \label{Ex2_J}\\
  r_{n,-1}^{X,\mathcal{D}}&=
  \frac{(b+4)(2g+2n-1)(2g+2n+5)(h+n-\tfrac32)_2}{8(a+2n-4)_2(a+2n-1)_4}\n
  &\quad\times\Bigl(
  b(b+9)\bigl(n(n+a-1)-(a-1)^2-1\bigr)\n
  &\qquad
  +2n(2a^2-4a+9)(n+a-1)+(a-1)^4-23(a-1)^2-14\Bigr),\n
  r_{n,0}^{X,\mathcal{D}}&=
  \frac{(b+4)(a-1)}{48(a+2n-3)_3(a+2n+1)_3}\Bigl(
     b^4(b+17)\bigl(6n(n+a)-(a-2)(a-3)\bigr)\n
  &\qquad
  -b^3\bigl(48n^3(n+2a)+48(a^2+a-14)n^2+48a(a-14)n\n
  &\qquad\qquad
  -(a-2)(a-3)(3a^2+3a-104)\bigr)\n
  &\quad
  -2b^2\bigl(264n^3(n+2a)+6(45a^2+42a-181)n^2+6a(a^2+42a-181)n\n
  &\qquad\qquad
  -(a-2)(a-3)(15a^2+12a-137)\bigr)\n
  &\qquad
  +3b\bigl(32n^5(n+3a)+16(6a^2+3a-43)n^4+32a(a^2+3a-43)n^3\n
  &\qquad\qquad
  -2(3a^4-36a^3+358a^2+316a-625)n^2\n
  &\qquad\qquad
  -2a(3a^4-12a^3+14a^2+316a-625)n\n
  &\qquad\qquad
  -(a-2)(a-3)(a^4+2a^3-32a^2-14a+99))\n
  &\qquad
  +3(a+2n-1)(a+2n+1)\bigl(24n^3(n+2a)+2(7a^2+22a-111)n^2\n
  &\qquad\qquad\qquad
  -2a(5a^2-22a+111)n-(a-2)(a-3)(4a^2+9a-33)\bigr)\Bigr),
  \nonumber
\end{align}
\underline{Ex.3} $\mathcal{D}=\{1^{\I},1^{\II}\}$ : 9-term recurrence relations
\begin{align}
  X(\eta)&=X_{\text{min}}(\eta)=
  -\frac{1}{64}\eta\bigl((b-2)b(b+2)\eta^3+4b^2(a-1)\eta^2+6b(a-1)^2\eta\n
  &\qquad\qquad\qquad\qquad\quad
  +4(a-3)(a-1)(a+1)\bigr),\n
  r_{n,4}^{X,\mathcal{D}}&=
  -\frac{(n+1)_4(b-2)b(b+2)(a+n)_4(2g+2n-3)(2h+2n-3)}
  {4(a+2n)_8(2g+2n+5)(2h+2n+5)},\n
  r_{n,-4}^{X,\mathcal{D}}&=
  -\frac{(b-2)b(b+2)}{64(a+2n-7)_8}(2g+2n-7)(g+n-\tfrac32)_2(2g+2n+3)\n
  &\qquad\times
  (2h+2n-7)(h+n-\tfrac32)_2(2h+2n+3),\n
  r_{n,3}^{X,\mathcal{D}}&=
  -\frac{(n+1)_3b^2(a-1)(a+n)_3(2g+2n-3)(2h+2n-3)}{2(a+2n-1)_7(a+2n+7)},\n
  r_{n,-3}^{X,\mathcal{D}}&=
  -\frac{b^2(a-1)(g+n-\tfrac52)_3(2g+2n+3)(h+n-\tfrac52)_3(2h+2n+3)}
  {2(a+2n-7)(a+2n-5)_7},\n
  r_{n,2}^{X,\mathcal{D}}&=
  \frac{(n+1)_2b(a+n)_2(2g+2n-3)(2h+2n-3)}
  {8(a+2n-2)_6(a+2n+5)_2(2g+2n+1)(2h+2n+1)}\n
  &\quad\times\Bigl(
  b^4\bigl(2n(n+a+2)-3(a-1)(a-2)\bigr)\n
  &\qquad
  -b^2\bigl(8n^3(n+2a+4)-2(7a^2-50a-15)n^2-2(a+2)(11a^2-34a+1)n\n
  &\qquad\qquad
  -3(a-1)(a-2)(2a^2+9a+11)\bigr)\n
  &\qquad
  -(a+2n-1)(a+2n+5)\bigl(4(3a^2-6a+1)n(n+a+2)\n
  &\qquad\qquad\qquad\qquad\qquad\qquad\qquad
  +3(a-2)(a+1)^2(a+2)\bigr)\Bigr),\n
  r_{n,-2}^{X,\mathcal{D}}&=
  \frac{b(2g+2n-3)(2g+2n+3)(2h+2n-3)(2h+2n+3)}
  {128(a+2n-6)_2(a+2n-3)_6}\n
  &\quad\times\Bigl(
  b^4\bigl(2n(n+a-2)-3a^2+5a-6\bigr)\n
  &\qquad
  -b^2\bigl(8n^4+16(a-2)n^3-2(7a^2-2a-15)n^2-2(a-2)(11a^2-18a+1)n\n
  &\qquad\qquad
  -6a^4+35a^3-68a^2+49a-66\bigr)\n
  &\qquad
  -(2n+a+1)(2n+a-5)\bigl(4(3a^2-6a+1)n(n+a-2)\n
  &\qquad\qquad\qquad\qquad\qquad\qquad\qquad
  +(a-3)(3a^3-9a^2+12a+4)\bigr)\Bigr),\\
  r_{n,1}^{X,\mathcal{D}}&=
  -\frac{(n+1)(a-1)(a+n)(2g+2n-3)(2g+2n+3)(2h+2n-3)(2h+2n+3)}
  {8(a+2n-3)_5(a+2n+3)_3}\n
  &\quad\times\Bigl(
  b^2\bigl(3n(n+a+1)-(a-2)(a-3)\bigr)+(a+1)(a-3)(a+2n-2)(a+2n+4)\Bigr),\n
  r_{n,-1}^{X,\mathcal{D}}&=
  -\frac{(a-1)(g+n-\tfrac32)_2(2g+2n+3)(h+n-\tfrac32)_2(2h+2n+3)}
  {2(a+2n-5)_3(a+2n-1)_5}\n
  &\quad\times\Bigl(
  b^2\bigl(3n(n+a-1)-a^2+2a-6\bigr)+(a-3)(a+1)(a+2n-4)(a+2n+2)\Bigr),\n
  r_{n,0}^{X,\mathcal{D}}&=
  -\frac{b}{64(a+2n-4)_4(a+2n+1)_4}\n
  &\quad\times\Bigl(
  b^6\bigl(6n^3(n+2a)-6(a^2-5a+5)n^2-6a(2a^2-5a+5)n+(a-4)_4\bigr)\n
  &\qquad
  -2b^4\bigl(24n^5(n+3a)+6(a^2+28a-9)n^4-12a(9a^2-28a+9)n^3\n
  &\qquad\qquad
  -2(38a^4-71a^3-17a^2+5a+117)n^2-2a(5a^4+13a^3-44a^2+5a+117)n\n
  &\qquad\qquad
  +(a-4)_4(2a^2+3a+11)\bigr)\n
  &\qquad
  +b^2\bigl(96n^7(n+4a)+48(a^2+26a-15)n^6-48a(25a^2-78a+45)n^5\n
  &\qquad\qquad
  -6(279a^4-616a^3+98a^2+376a-417)n^4\n
  &\qquad\qquad
  -12a(75a^4-96a^3-202a^2+376a-417)n^3\n
  &\qquad\qquad
  -2(87a^6+153a^5-1139a^4+1262a^3-931a^2-1031a+2775)n^2\n
  &\qquad\qquad
  +2a(6a^6-129a^5+353a^4-134a^3-320a^2+1031a-2775)n\n
  &\qquad\qquad
  +(a-4)_4(6a^4+18a^3+37a^2+114a+153)\bigr)\n
  &\qquad
  +2(a+2n-3)(a+2n+3)\bigl(
  48(2a^2-4a+1)n^5(n+3a)\n
  &\qquad\qquad\qquad
  +4(76a^4-124a^3-73a^2+124a-39)n^4\n
  &\qquad\qquad\qquad
  +8a(16a^4-4a^3-103a^2+124a-39)n^3\n
  &\qquad\qquad\qquad
  +2(3a^6+78a^5-274a^4+142a^3+385a^2-544a-114)n^2\n
  &\qquad\qquad\qquad
  -2a(5a^6-38a^5+56a^4+106a^3-463a^2+544a+114)n\n
  &\qquad\qquad\qquad
  -(a-4)(a-2)_2(a+1)_2(2a^3-3a^2-2a+21)\bigr)\Bigr).
  \nonumber
\end{align}
Note that we have equivalences \cite{equiv_miop},
\begin{align}
  P_{\{1^{\I},2^{\I}\},n}(\eta;g,h)
  &=-\frac{(g-h+4)(h+n-\tfrac52)_2}{4(g+n+\tfrac12)}
  P_{\{2^{\II}\},n}(\eta;g+3,h-3),\\
  P_{\{1^{\I},1^{\II}\},n}(\eta;g,h)
  &=\frac{3(g+n-\frac32)}{(g-h+1)(h+n+\frac12)}
  P_{\{1^{\I},3^{\I}\},n}(\eta;g-2,h+2).
\end{align}
Recurrence relations for $P_{\{1^{\I}\},n}(\eta;g,g)$ and 
$P_{\{2^{\II}\},n}(\eta;g,g)$ were given in \cite{mt}.

\subsection{Multi-indexed Wilson polynomials}
\label{sec:B_W}

We set $b_1=a_1+a_2+a_3+a_4$, $\sigma_1=a_1+a_2$, $\sigma_2=a_1a_2$,
$\sigma'_1=a_3+a_4$ and $\sigma'_2=a_3a_4$.\\
\noindent
\underline{Ex.1} $\mathcal{D}=\{1^{\I}\}$ : 5-term recurrence relations
\begin{align}
  X(\eta)&=X_{\text{min}}(\eta)=\tfrac14\eta\bigl(2(\sigma_1-\sigma'_1-2)\eta
  +4(\sigma_2\sigma'_1-\sigma_1\sigma'_2-\sigma_1\sigma'_1+2\sigma'_2)
  +\sigma_1+3\sigma'_1-2\bigr),\n
  r_{n,2}^{X,\mathcal{D}}&=
  \frac{(\sigma_1-\sigma'_1-2)(b_1+n-1)_2(\sigma_1+n-2)}
  {2(b_1+2n-1)_4(\sigma_1+n)},\n
  r_{n,-2}^{X,\mathcal{D}}&=
  \frac{n(n-1)(\sigma_1-\sigma'_1-2)}{2(b_1+2n-4)_4}
  (\sigma_1+n-2)_2(\sigma'_1+n-2)(\sigma'_1+n+1)\n
  &\quad\times
  \prod_{i=1}^2\prod_{j=3}^4(a_i+a_j+n-2)_2\,,\n
  r_{n,1}^{X,\mathcal{D}}&=
  -\frac{2(b_1+n-1)(\sigma_1+n-2)(\sigma'_1+n+1)}
  {(b_1+2n-2)_3(b_1+2n+2)}\n
  &\quad\times
  \bigl((\sigma_1-\sigma'_1-2)n(n+b_1)
  -(b_1-2)(\sigma'_1-\sigma_2+\sigma'_2+1)\bigr),\\
  r_{n,-1}^{X,\mathcal{D}}&=
  \frac{2n(\sigma_1+n-2)_2(\sigma'_1+n-1)(\sigma'_1+n+1)}
  {(b_1+2n-4)(b_1+2n-2)_3}
  \prod_{i=1}^2\prod_{j=3}^4(a_i+a_j+n-1)\n
  &\quad\times
  \bigl((2-\sigma_1+\sigma'_1)n(n+b_1-2)+(\sigma_1-2)b_1
  -(\sigma_2-\sigma'_2)(b_1-2)\bigr),\n
  r_{n,0}^{X,\mathcal{D}}&=
  \frac{1}{32(b_1+2n-3)_2(b_1+2n)_2}A.
  \nonumber
\end{align}
Here $A$ is a polynomial of degree $8$ in $n$, whose coefficients are
polynomials in $\sigma_1$, $\sigma_2$, $\sigma'_1$ and $\sigma'_2$.
Since $A$ has a lengthy expression, we do not write down it here and
put it on the web page \cite{wp}.

\subsection{Multi-indexed Askey-Wilson polynomials}
\label{sec:B_AW}

We set $b_4=a_1a_2a_3a_4$, $\sigma_1=a_1+a_2$, $\sigma_2=a_1a_2$,
$\sigma'_1=a_3+a_4$ and $\sigma'_2=a_3a_4$.\\
\noindent
\underline{Ex.1} $\mathcal{D}=\{1^{\I}\}$ : 5-term recurrence relations
\begin{align}
  X(\eta)&=X_{\text{min}}(\eta)=\frac{\eta}{(1+q)\sigma_1}
  \Bigl(2q^{\frac12}(\sigma_2-\sigma'_2q^2)\eta
  -(1+q)\bigl(\sigma_1(1-\sigma'_2)q+\sigma'_1(\sigma_2-q^2)\bigr)\Bigr),\n
  r_{n,2}^{X,\mathcal{D}}&=
  \frac{q^{\frac32}(1-\sigma_2^{-1}\sigma'_2q^2)(b_4q^{n-1};q)_2
  (1-\sigma_2q^{n-2})}
  {2(1+q)(b_4q^{2n-1};q)_4(1-\sigma_2q^n)},\n
  r_{n,-2}^{X,\mathcal{D}}&=
  \frac{(q^{n-1};q)_2(1-\sigma_2^{-1}\sigma'_2q^2)}
  {2(1+q)q^{\frac12}(b_4q^{2n-4};q)_4}
  (\sigma_2q^{n-2};q)_2(1-\sigma'_2q^{n-2})(1-\sigma'_2q^{n+1})
  \prod_{i=1}^2\prod_{j=3}^4(a_ia_jq^{n-2};q)_2,\n
  r_{n,1}^{X,\mathcal{D}}&=
  -\frac{(1-b_4q^{n-1})(1-\sigma_2q^{n-2})(1-\sigma'_2q^{n+1})}
  {2q^{\frac12}\sigma_2(b_4q^{2n-2};q)_3(1-b_4q^{2n+2})}\n
  &\quad\times
  \Bigl(q(b_4q^{2n}+1)
  \bigl(q\sigma_1(1-\sigma'_2)+\sigma'_1(\sigma_2-q^2)\bigr)\n
  &\qquad\quad
  -(1+q^2)q^n\bigl(\sigma_1\sigma'_2(\sigma_2-q^2)
  +\sigma'_1\sigma_2q(1-\sigma'_2)\bigr)\Bigr),\\
  r_{n,-1}^{X,\mathcal{D}}&=
  -\frac{(1-q^n)(\sigma_2q^{n-2};q)_2
  (1-\sigma'_2q^{n-1})(1-\sigma'_2q^{n+1})}
  {2q^{\frac52}\sigma_2(1-b_4q^{2n-4})(b_4q^{2n-2};q)_3}
  \prod_{i=1}^2\prod_{j=3}^4(1-a_ia_jq^{n-1})\n
  &\quad\times
  \Bigl((b_4q^{2n}+q^2)
  \bigl(q\sigma_1(1-\sigma'_2)+\sigma'_1(\sigma_2-q^2)\bigr)\n
  &\qquad\quad
  -(1+q^2)q^n\bigl(\sigma_1\sigma'_2(\sigma_2-q^2)
  +\sigma'_1\sigma_2q(1-\sigma'_2)\bigr)\Bigr),\n
  r_{n,0}^{X,\mathcal{D}}&=
  \frac{1}{2q^{\frac{11}{2}}\sigma_2(1+q)(b_4q^{2n-3};q)_2(b_4q^{2n};q)_2}A,
  \nonumber
\end{align}
Here $A$ is a polynomial of degree $8$ in $q^n$, whose coefficients are
polynomials in $\sigma_1$, $\sigma_2$, $\sigma'_1$ and $\sigma'_2$.
Since $A$ has a lengthy expression, we put it on the web page \cite{wp}.


\end{document}